\documentclass[useAMS,usenatbib]{mn2e}

\usepackage{array}
\usepackage{graphicx}
\usepackage{amsmath}
\usepackage[version=3]{mhchem}
\usepackage{textcomp}

\usepackage{txfonts}
\def\nh{n$_\textrm{H}$}

\def\cmt{cm$^{-3}$}
\def\cmd{cm$^{-2}$}

\def\hh{H$_2$}
\def\hho{H$_2$O}
\def\nhh{NH$_2$}

\def\coo{CO$_2$}

\def\chhhh{CH$_4$}
\def\hhco{H$_2$CO}
\def\chhho{CH$_3$O}

\def\chhhoh{CH$_3$OH}

\newcommand{\tabitem}{~~\llap{\textbullet}~~}

\title[Modeling Complex Organic Molecules in dense regions]{Modeling Complex Organic Molecules in dense regions: Eley-Rideal and complex induced reaction}
\author[M. Ruaud]{M. Ruaud$^{1}$$^{2}$\thanks{E-mail:
ruaud@obs.u-bordeaux1.fr}, J. C. Loison $^{3}$$^{4}$\thanks{E-mail:
jean-christophe.loison@u-bordeaux.fr}, K. M. Hickson $^{3}$$^{4}$, P. Gratier $^{1}$$^{2}$, F. Hersant $^{1}$$^{2}$, V. Wakelam $^{1}$$^{2}$\\
$^{1}$Univ. Bordeaux, LAB, UMR 5804, F-33270, Floirac, France\\
$^{2}$CNRS, LAB, UMR 5804, F-33270, Floirac, France\\
$^{3}$Univ. Bordeaux, ISM, UMR 5255, F-33400 Talence, France\\
$^{4}$CNRS, ISM, UMR 5255, F-33400 Talence, France}
\begin{document}

\date{Accepted 2014 December 18. Received 2014 December 18; in original form 2014 November 17}

\pagerange{\pageref{firstpage}--\pageref{lastpage}} \pubyear{2014}

\maketitle

\label{firstpage}

\begin{abstract}
Recent observations have revealed the existence of Complex Organic Molecules (COMs) in cold dense cores and prestellar cores. The presence of these molecules in such cold conditions is not well understood and remains a matter of debate since the previously proposed "warm-up" scenario cannot explain these observations. In this article, we study the effect of Eley-Rideal and complex induced reaction mechanisms of gas-phase carbon atoms with the main ice components of dust grains on the formation of COMs in cold and dense regions. Based on recent experiments we use a low value for the chemical desorption efficiency (which was previously invoked to explain the observed COM abundances). We show that our introduced mechanisms are efficient enough to produce a large amount of complex organic molecules in the gas-phase at temperatures as low as 10K. 
\end{abstract}

\begin{keywords}
astrochemistry -- ISM: abundances -- ISM: clouds -- molecules
\end{keywords}

\maketitle

\section{Introduction}
Complex Organic Molecules (COM here after), such as  CH$_3$OH, CH$_3$CHO, HCOOCH$_3$ and CH$_3$OCH$_3$ \citep[see][for a review]{Herbst09}, have long been observed in hot cores or hot corinos of warm star-forming regions such as SgB2 \citep{Cummins86}, OMC-1 \citep{Blake87} or IRAS 16293-2422 \citep{Bottinelli04}. Although some of these molecules such as CH$_3$OH and CH$_3$CHO were also observed in cold environments like the TMC-1 or L134N dark clouds \citep{Matthews85,Friberg88}, a search for more complex COMs dimethyl ether yielded only upper limits \citep{Friberg88}. In the B1-b dense core, \citet{Oberg10} reported detections of CH$_3$OH, CH$_3$CHO, HCOOCH$_3$ and tentatively detected CH$_3$OCH$_3$. Recently, \citet{Cernicharo12} reported the discovery of the methoxy radical (CH$_3$O) in this source. Furthermore, observations conducted by \citet{Bacmann12} have revealed the existence of a variety of COMs in the cold prestellar core L1689b and more recently toward the prestellar core L1544 \citep{Vastel14}. From these detections, new challenges arise for current chemical models. Indeed, in star forming regions, these COMs are thought to be formed via recombination of radicals, at temperatures around 30K, on ices during the warm-up period which are then released into the gas-phase at temperatures above 100K  \citep{Garrod06}. In cold cores which are characterized by a very low temperature ($\sim$10K), this mechanism cannot explain the observed abundance of such molecules.

Recently, \citet{Vasyunin13}, have proposed that these complex species are formed via gas phase ion-molecular and neutral-neutral chemistry from precursors such as formaldehyde and methanol. In this scenario, these precursors are formed on icy grain surfaces thanks to the surface mobility of atoms such as hydrogen, oxygen and nitrogen. These precursors are then released to the gas-phase via efficient reactive desorption (i.e $\sim10\%$ of the newly formed species desorb) \citep{Garrod07}. The proposed scenario leads to reasonable results at temperature as low as 10K for a range of molecules. However the reactive desorption mechanism is not a fully understood process. Recent experiments conducted by \citet{Minissale14} showed that this process may not be as efficient as previously thought (see discussion in Section \ref{gas_grain_code}). It is important to note that thermal and cosmic ray induced desorption mechanisms are unable to reproduce the observed COM abundances in their simulations.

In this article, we propose an alternative/complementary scenario, in which Eley-Rideal and the complex induced reaction mechanisms on grain surfaces modify the chemical reactivity and can be efficient under some conditions to reproduce the observed COM abundances in cold and dense regions.
Indeed, usual gas-grain models consider that atoms and molecules present in the gas-phase may stick on interstellar dust grains through collisions. These species are only weakly bound to the surface (physisorbed), considering in most cases a sticking probability equal to 1 \citep{Hasegawa92}. These species can then react through the Langmuir-Hinshelwood mechanism (i.e. diffusion process). 

In most astrophysical models, Eley-Rideal (atom or molecule reacting directly with adsorbed species) mechanisms are neglected considering the low density of reactive species on grains. This may be not the case for carbon atom collisions with interstellar ices as carbon atoms show high reactivity with most of the molecules adsorbed on grain surfaces and are present in the gas phase at high abundances.

Another critical point is the strength of the adsorption itself. In current astrochemical models the adsorption energy of a carbon atom on ice is taken to be 800 K \citep{Hama13}. This value has neither been measured nor calculated and could be significantly underestimated as ab-initio calculations lead to a binding energy of the C...\hho~ van der Waals complex equal to 3600 K \citep{Schreiner06, Hwang99, Ozkan11} and to a binding energy of the C...\chhhoh~ van der Waals complex equal to 8400 K \citep{Dede12}. Then, even if carbon atoms have been shown to react quickly with methanol in the gas phase \citep{Shannon13}, the strength of the C...\chhhoh~ complex may prevent reaction if the energy dissipation is efficient enough to stabilize the complex. 

This paper is organized as follows. In Section 2, we describe our chemical model and we introduce the basics of the considered mechanisms. In Section 3, we present the effect of Eley-Rideal and complex induced reaction mechanisms on our models and on the formation of COMs. In Section 4, we compare our modeling results with the observed COM abundances in B1-b and L1689b cold dense cores and L1544 prestellar core, showing that these processes play a significant role in reproducing the observed abundances. Finally, Section 5 contains a general discussion and a summary of our work.

\section{Chemical model}

\subsection{The gas-grain code} \label{gas_grain_code}
For this work, we used the full gas-grain chemical model {\sevensize NAUTILUS} described in \citet{Semenov10} and \citet{Reboussin14}. This model is adapted from the original gas-grain model of \citet{Hasegawa92} and its subsequent evolutions made over the years at Ohio State University. In this model, the abundance of each species is computed by solving a set of rate equations for gas-phase and grain-surface chemistries. Gas-phase and grain-surface chemistries are connected via accretion and desorption. Because of the low grain temperature expected in cold dense cores, we only consider physisorption. Desorption can be thermal, induced by stochastic cosmic ray heating or via the exothermicity of surface reactions. The rate at which surface bound species thermally desorb follows a Boltzmann law at the dust temperature \citep{Hasegawa92}. The cosmic ray induced desorption is treated following \citet{Hasegawa93} and the corresponding rate is given by:
\begin{equation}
   k_{crd}(i) = f(70\textrm{K})k_{des}(i,70\textrm{K})
\end{equation}
where $f$(70K) is the fraction of the time spent by grains at 70K which is estimated to be $\sim3.16\times10^{-19}$ for a cosmic ionization rate $\zeta_{\textrm{H}_2}=1.3\times10^{-17}$ s$^{-1}$ and $k_{des}(i,70\textrm{K})$ the rate of thermal desorption for a grain at T$_\textrm{dust}$=70K. The desorption by exothermicity of surface reactions is treated following \citet{Garrod07}. In this scenario, for each surface reaction that leads to a single product, we consider that a part of the energy released during the reaction could contribute to break the surface-molecule bond. The fraction $f$ of reactions resulting in desorption is calculated by modeling the competition between the rate of desorption and the rate of energy lost to the grain:
\begin{equation}
   f = \frac{\nu P}{\nu_s + \nu P} = \frac{aP}{1+aP}
\end{equation}
where $a=\nu/\nu_s$, the ratio of the surface-molecule bond frequency to the energy at which energy is lost to the grain surface. $P$ gives the probability of desorption and follows the Rice-Ramsperger-Kessel (RRK) theory \citep[see][and references therein for more information]{Garrod07}. The value $a$ is not well constrained but could be in the range of $\sim$0.0 (no chemical desorption) to $\sim$0.1 (high efficiency). However, recents experiments conducted by \citet{Minissale14} on the efficiency of the reactive desorption of the oxygen system (i.e. system with two open channels leading to O$_2$ and O$_3$) on oxidized graphite showed that this process is highly dependent on the surface coverage and linearly decreases with increasing surface coverage. They explain this phenomenon saying that the presence of a pre-adsorbed species possibly enhances the probability for the newly formed excited molecules to dissipate their excess energy, presumably making the chemical desorption process inefficient when the grain coverage is important. 

Therefore, under these conditions, we choose to use a moderate value for the reactive desorption efficiency. We considered a value of $a=0.01$, which is equivalent to consider that $\sim 1\%$ of the newly formed species desorb at formation and $\sim 99\%$ remain on the grain surface \citep[see][]{Garrod07}.

The surface chemistry is treated assuming that the ice mantle surrounding the interstellar grain core can be represented as an isotropic lattice with periodic potential. The width of barrier between two adjacent  binding sites is set to $1 \AA$ and the height E$_b$ of the barrier against diffusion is assumed to be $0.5\times$ E$_D$ \citep{Garrod06}, where E$_D$ is the desorption energy of the species. Atoms and molecules can diffuse on the grain surface by thermal hopping and then react with each other, according to \citet{Hasegawa92} (we do not take into account diffusion by quantum tunneling in this study).

We also assume the "encounter desorption" process proposed by \citet{Hincelin14} for grain surface molecular hydrogen. The determination of the \hh~coverage on a dust grain is critical due to the very different binding energies involved depending on whether \hh~ adsorbs on water substrate (E$_D$=440K) or on \hh~substrate (E$_D$=23K)\citep{Cuppen07}. Since \hh~is the most abundant species in the gas phase and because s-\hho~ (water molecule on surface) is thought to be the main component of the ice mantle, the use of a single binding energy for \hh~can lead to very different results depending on the assumptions made. Indeed, considering that \hh~adsorbate on a water substrate can lead, at high density, to the formation of several \hh~monolayers on grain surfaces, is physically unacceptable considering the very low binding energy of \hh~on itself.  In contrast, considering the binding energy of \hh~on itself prevents the adsorption of \hh~onto grain surfaces, which is not satisfactory at low temperature. The proposed approach by Hincelin et al. (2014) is based on the facile desorption of molecular hydrogen when it is adsorbed on an \hh~substrate and deals with these two binding energies. This method leads to reasonable results with a microscopic Monte Carlo stochastic method.

Finally, grains are considered to be spherical with a $0.1\mu$m radius, a 3 g.cm$^{-3}$ density and $\sim 10^6$ surface sites, all chemically active. The dust to gas mass ratio is set to 0.01. 

\subsection{The new chemical network}
For this study, to calculate the various desorption rates, and the diffusive reaction rates, we used the binding energies E$_D$ from \citet{Garrod06}. We adopted the gas-phase network \textit{kida.uva.2011} with updates from \citet{Loison12}, \citet{Wakelam13} and \citet{Loison2014}. The grain network is based on \citet{Garrod07}. The grain surface network has been modified in order to include the new surface mechanisms (formation of van der Waals complexes and low temperature Eley-Rideal mechanisms). Several new species have been added to the model (see Table \ref{binding_energies}; adopted binding energies are also reported with their respective references) together with the reactions involving those species in the gas phase. The new mechanisms are described in section~\ref{chem} and the new reactions are listed in Tables \ref{grain_chemistry} and \ref{gas_chemistry} of Appendix \ref{network}. The full network is available at \verb+http://kida.obs.u-bordeaux1.fr/models+. 

\subsubsection{Chemistry}\label{chem}

\begin{figure*} 
   \includegraphics[width= 14cm,trim = 0cm 1.2cm 0cm 7cm, clip]{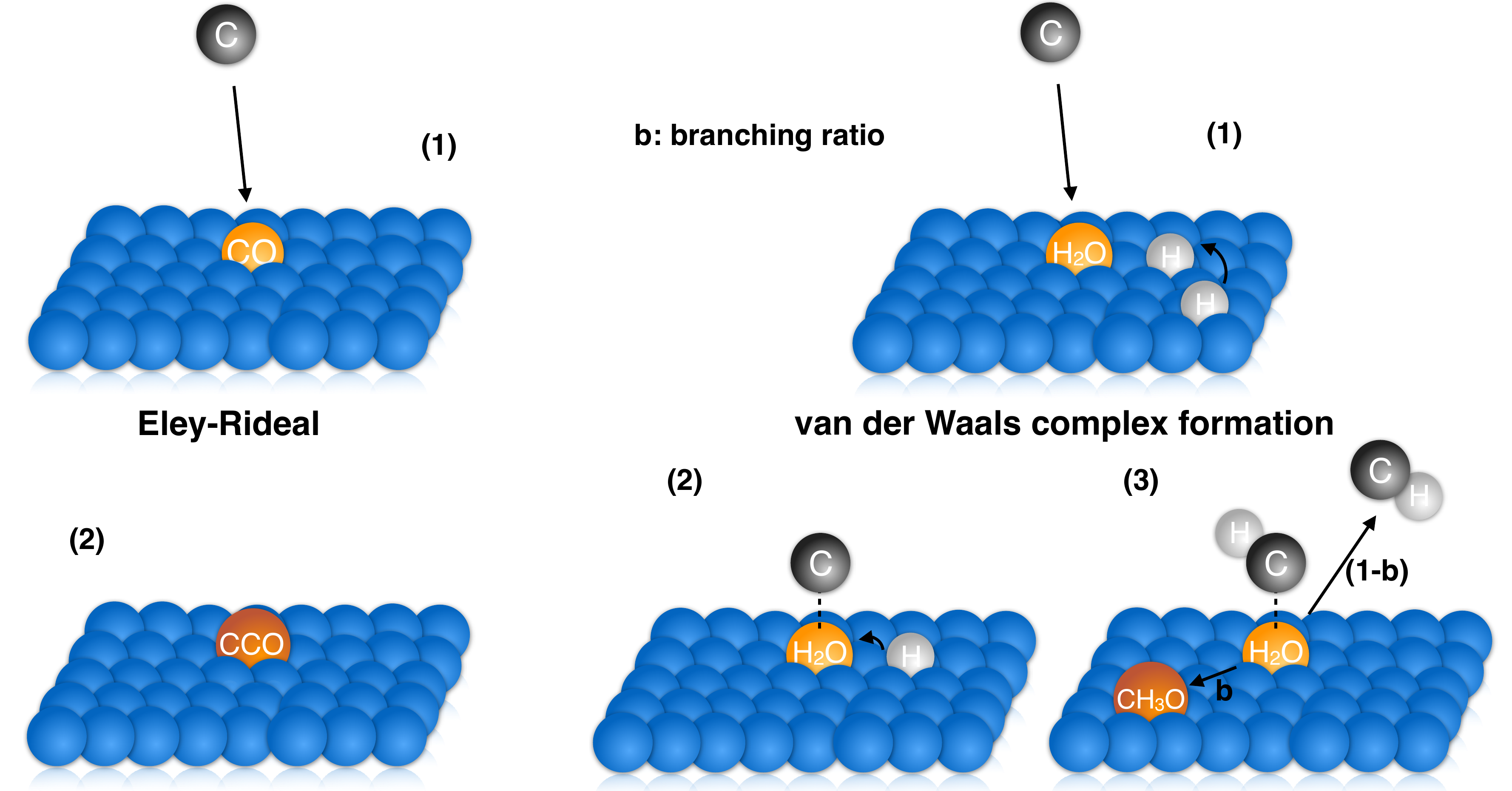}
   \caption{Sketch illustrating the Eley-Rideal and complex induced reaction mechanisms on grain surfaces.} 
   \label{sketch}
\end{figure*}

\begin{table*}
   \begin{center}
   \caption{Summary of grain surface processes considered in this study}
   \label{summary}
   \begin{tabular}{@{}lcll}
   \hline
   \hline
   \multicolumn{3}{l}{Reaction$^\dagger$}								& Assumptions and comments\\
   \hline
   C + s-X					&$\rightarrow$ & s-CX					& Eley-Rideal mechanism when there is no barrier for reaction: s-H$_2$, s-CO and s-H$_2$CO\\
   						&$\rightarrow$ & s-C...X		 			& Complex induced reaction mechanism otherwise: s-H$_2$O, s-CO$_2$, s-CH$_3$OH, s-CH$_4$ and s-NH$_3$\\
						&			&						&\\
   						&			&						&Due to the large value of CH internal energy and because carbon atom complexes are characterized by short van\\
						&			&						& der Waals distances (i.e. they should react easily), we consider that: \\
   s-C...X + s-H				&$\rightarrow$ & s-CHX					& ~~\tabitem most leads to more complex molecules (within a part will desorb due to reactive desorption),\\
						&$\rightarrow$ & s-CH...X					& ~~\tabitem a small fraction remain on the form of s-CH...X (i.e. 1\% in most cases),\\
						&$\rightarrow$ & CH + s-X				& ~~\tabitem a small fraction of s-CH radicals desorb (i.e. 1\%).\\ 
						&			&						&\\
						&			&						& The O...CO case:\\
   s-O...CO + s-H 			&$\rightarrow$ & s-HOCO					& ~~\tabitem 19\% lead to s-HOCO formation (within a part will desorb due to reactive desorption),\\ 
    						&$\rightarrow$ & s-OH + s-CO				& ~~\tabitem 20\% of s-OH relaxation (much larger than for carbon atoms due to the larger van der Waals distance),\\
						&$\rightarrow$ & s-H + s-CO$_2$			& ~~\tabitem 60\% of s-CO$_2$ formation,\\
						&$\rightarrow$ & OH + s-CO				& ~~\tabitem 1\% of s-OH desorption.\\
   \hline
   \end{tabular}
   \end{center}
\end{table*}

In this study, we consider that when gas-phase carbon atoms land on grain surfaces they can either react directly (i.e. Eley-Rideal mechanism, C + s-X $\rightarrow$ s-CX) or form van der Waals complexes (C + s-X $\rightarrow$ s-C...X) with the main constituent molecules present on interstellar grains (i.e. s-X = s-H$_2$O, s-CO, s-NH$_3$, s-CO$_2$, s-CH$_4$, s-H$_2$CO and s-CH$_3$OH, see Fig. \ref{sketch}) or simply be physisorbed. 

We consider that when there is no barrier for reaction, the C + s-X reaction occurs via the Eley-Rideal mechanism. This is the case for reactions C + H$_2$, C + CO and C + H$_2$CO which have no barrier \citep{Harding83,Harding93,Husain71,Husain99}. On the other hand, for reactions such as C + s-H$_2$O, C + s-CO$_2$, C + s-NH$_3$, C + s-CH$_4$ and C + s-CH$_3$OH, we consider that atomic carbon will preferentially form a complex due to the presence of a deep van der Waals complex well. Consequently they will not react except through tunneling assuming that the height of the barrier is relatively unaffected by the presence of ice. This latter approximation is rather uncertain \citep{Bromley14} but has been shown to be valid for some systems using theoretical calculations: O + CO \citep{Talbi06}, H + HOCCOH \citep{Woods13}, H + H$_2$CO \citep{Goumans11}. Moreover, as the van der Waals energies are large, we consider that the s-C...X complexes are unable to move on the surface (see Table \ref{binding_energies}). 

When a van der Waals complex is formed, atomic carbon will react with the highly mobile H atoms through reactions: s-C...X + s-H. The products of these reactions is energized s-CH...X radicals (e.g. internal energy for the CH...H$_2$O complex of $\sim$ 358 kJ/mol corresponding to the value of the exothermicity of the C + H $\rightarrow$ CH reaction added to the difference of C...\hho~ \citep{Schreiner06, Hwang99, Ozkan11} and CH...\hho~ \citep{Bergeat09} van der Waals bond strengths). Considering the large value of the CH internal energy, energy dissipation is unlikely to be efficient enough to stabilize the CH...ice complex (see Table \ref{binding_energies}). We therefore consider that a small fraction of the CH radicals will desorb (i.e. 1\%), another small fraction (i.e. 1\%) will remain in the form of the CH...X complex and that most will react with the species X leading to more complex molecules. The reactions s-C...X + s-H $\rightarrow$ s-CH...X $\rightarrow$ s-Y are considered to have no barrier similary to the gas-phase for most of the CH reactions \citep{Hickson13,Canosa97,Bocherel96}. The different reactions and products (with branching ratios) are given in Table \ref{grain_chemistry} of Appendix \ref{network}. 

The products of surface reactions are usually considered to be adduct formation and the bimolecular exit channels are neglected given the fast energy dissipation (equivalent to a gas phase reaction with infinite pressure). For example, the s-CH + s-\chhhh~ reaction is assumed to lead only to s-C$_2$H$_5$ and not to s-C$_2$H$_4$ + s-H. This approximation is almost systematically used in current astrophysical models of grain chemistry and has yet to be verified by theoretical calculations. Nevertheless, it is likely to be a poor assumption for reactions where the bimolecular exit channel is very exothermic involving a low exit barrier. A good example is the s-CH + s-\coo~ reaction leading to s-HCOCO adduct formation which may partly fall apart to give s-HCO + s-CO. Even if s-CO and s-HCO remain relatively close together on the ice, they will lose their internal and kinetic energies through interaction with the surface preventing reformation of the s-HCOCO adduct which probably presents a notable barrier. The reactions of atomic species such as N, O with H on interstellar ices will also result in the formation of other highly energetic hydride species reagents. In the case of atomic nitrogen, the resulting s-NH and s-\nhh~ radicals could then possess a significantly increased reactivity with respect to their ground state counterparts. However, as s-NH and s-\nhh~ both show large to very large barriers for reactions with s-\hho, s-CO, s-\coo, s-\chhhoh, s-\chhhh~ and s-\hhco, we currently neglect these reactions \citep{Rohrig94,Cohen91}.

The case of energized OH (formed by the O + H addition) is different. OH radicals show a high reactivity with \chhhoh~ \citep{Jimenez03, Xu07,Shannon13} and \hhco~ \citep{Yetter89, Xu06} leading to H atom abstraction and then to \chhho~ and HCO radicals and water. As s-\chhho~ and s-HCO radicals will mainly react with the highly mobile s-H atoms on the surface, reforming s-\chhhoh~ and s-\hhco, the main global reaction is s-OH + s-H $\rightarrow$ s-\hho. The only specific case is when an O atom is bound to a CO molecule on the ice surface, s-O...CO, as energized OH formed by the s-O...CO + s-H reaction may react with s-CO to form s-HOCO. Indeed the OH + CO $\rightarrow$ HOCO reaction shows a very small barrier \citep{Fulle96, Joshi06, Nguyen12}. The energized HOCO adduct may then fall apart further to yield H + \coo~ as shown in a recent \hho-CO ice photodissociation experiment \citep{Arasa13}. We introduce the s-O...CO van der Waals complex and HOCO in the model. We performed DFT calculations showing that HOCO reacts with H atoms without a barrier, producing formic acid (HCOOH) or \hh~ + \coo~ through direct H atom abstraction \citep[in good agreement with][]{Yu08,Dibble10}. In their \hho-CO ice photodissociation study, \citet{Arasa13} calculated that most of the OH is relaxed, $3\%$ of OH radicals lead to HOCO formation and only $0.036\%$ lead to \coo~ + H formation. However in their case the OH formed by \hho~ photodissociation had an average OH internal energy estimated to be equal to 68 kJ/mol, much smaller than in the case of OH formation from the O + H addition reaction (430 kJ/mol). We consider that energized OH from the s-O...CO + s-H addition leads to $1\%$ of OH desorption, $20\%$ of OH relaxation, $19\%$ to s-HOCO formation and $60\%$ to s-H + s-\coo~ formation. For the s-H + s-HOCO reaction, we consider that $10\%$ of reactive events lead to s-HCOOH formation, $19\%$ to s-\hho~+ s-CO with the other $70\%$ leading to s-\hh~ + s-\coo~ formation. As already mentioned, in the case of energized CH, we assume only $1\%$ of relaxation, compared to a value of $20\%$ for the equivalent energized OH formation process. The reason for this assumption is that the van der Waals complexes of atomic carbon are generally much stronger than those of oxygen atoms (O...CO stabilization is calculated equal to 258 K \citep{Goumans10}) so that carbon atom complexes are characterized by shorter van der Waals distances. As a result, energized CH radicals are produced much closer to the molecule and should react more easily.

 All the introduced processes and assumptions made are summarized in Table \ref{summary} and illustrated by the sketch of the Fig. \ref{sketch}.

\subsubsection{Formalism in the gas-grain code}

\begin{table}
   \begin{center}
   \caption{Desorption energies of added species.}
   \label{binding_energies}
   \begin{tabular}{@{}lrl}
   \hline
   \hline
   Species			&	E$_D$ (K) & References\\
   \hline
   C...H$_2$O			&	3600		&	\citet{Schreiner06}\\
   C...NH$_3$			&	14400	&	Ab-initio calculations (this work)\\
   C...CO$_2$			&	1200		&	M06-2X/cc-pVTZ (this work)\\
   C...CH$_4$			&	400		&	\citet{Kim03}  (CCSD(T)/6-311+G(3df,2p))\\
   C...CH$_3$OH		&	8400		&	\citet{Dede12}\\
   O...CO				&	258		& \citet{Goumans10}	\\
   CH...H$_2$O			&	5800		&	\citet{Bergeat09}\\
   CH...NH$_3$			&	13470	&	\citet{Blitz12}\\
   CH...CO$_2$			&	1800		&	M06-2X/cc-pVTZ (this work)\\
   CH...CH$_3$OH		&	5800		&	=CH...H$_2$O\\
   CH$_2$...CO$_2$		&	1050		&	estimation\\
   CH$_3$...CO$_2$		&	1175		&	estimation\\
   CH$_4$...CO$_2$		&	1300		&	estimation\\
   CH$_3$CO			&	2650		&	\citet{Garrod06} (=CH$_2$CHO)\\
   CH$_3$OCH$_2$		&	3500		&	estimation\\
   CH$_2$NH$_2$		&	5530		&	\citet{Hama13} (=CH3OH)\\
   CH$_2$NH			&	5530		&	\citet{Hama13} (=CH3OH)\\
   HOCO				&	2000		&	\citet{Hama13} (/H$_2$CO)\\
   HCOCO				&	2050		&	\citet{Hama13} (=H$_2$CO) \\
   HCOCHO			&	2050		&  	\citet{Hama13} (=H$_2$CO)\\
   CH$_3$O			&	5084		&      \citet{Garrod06} (=CH$_2$OH)\\
   \hline
   \end{tabular}
   \end{center}   
\end{table}

The Eley-Rideal and complexation mechanisms are computed assuming that for an incident species $i$ and an adsorbed species $j$ on the grain, the corresponding rate is given by:
\begin{equation} \label{eq:eley_rideal}
  R_{ij} = \eta_j \sigma_d \langle v(i) \rangle n(i) n_d ~~ [\textrm{cm}^{-3} \textrm{s}^{-1}],
\end{equation}
where $\eta_j = n_s(j)/ \sum_{k} n_s(k)$ represents the average density of the molecule on the surface, $\sigma_d$ the cross section of the grain, $\langle v(i) \rangle$ the thermal velocity of the incident species $i$, $n(i)$ its abundance and $n_d$ the number density of grains. $n_s(k)$ refers to the surface abundance of the species $k$. 

\section{Modeling results}

In the following sections, we present the results obtained with the new mechanisms introduced. For the simulations, we consider typical cold dense core conditions \citep{Vasyunin13} for our nominal model, i.e \nh=$1.0\times10^5$ \cmt, T$_\textrm{gas} $= T$_\textrm{dust}$=10K, A$_V$=10 and $\zeta_{\textrm{H}_2}$ =  $1.3\times10^{-17}$ s$^{-1}$. Initial input parameters and initial abundances are summarized in Tables \ref{cold_dense_cloud_model} and \ref{initial_abundances}. We first compare the nominal model (hereafter Model A) with Model B, which takes into account the mechanisms described in the previous section (i.e., carbon and oxygen complex induced reactions and the Eley-Rideal mechanism). We also considered an additional model in which we switched off the grain surface reaction s-H + s-CO $\rightarrow$ s-HCO in order to test our mechanism on the formation of methanol against the classical reaction scheme (Model C). We then performed a sensitivity analysis of our mechanism on the grain temperature, the cosmic ray ionization rate and the density.

\begin{table}
   \begin{center}
   \caption{Standard cold dense cloud model}
   \label{cold_dense_cloud_model}
   \begin{tabular}{@{}lcc}
   \hline
   \hline
   Parameter				& Value	\\
   \hline
   T \dotfill			 		& 10 K \\							
   n$_\textrm{H}$\dotfill		& $1\times10^5$ \cmt \\
   A$_V$\dotfill				& 10 \\
   $\zeta_{\textrm{H}_2}$\dotfill	& $1.3\times10^{-17}$ s$^{-1}$ \\
   Initial abundances\dotfill 	& Atomic (except for H) \\
   \hline
   \end{tabular}
   \end{center}  
\end{table}

\begin{table}
   \begin{center}
   \caption{Initial abundances used in our models.}
   \label{initial_abundances}
   \begin{tabular}{@{}lr}
   \hline
   \hline
   Element			&	$n_i/n_\textrm{H}$$^\dagger$   \\
   \hline
   H$_2$ 			&	0.5		\\
   He				&	0.09 $^\textrm{a}$		\\	
   N				&	6.2(-5) $^\textrm{b}$		\\	
   O				&	2.4(-4) $^\textrm{c}$		\\	      
   C$^+$			&	1.7(-4) $^\textrm{b}$		\\	   
   S$^+$			&	8.0(-9) $^\textrm{d}$		\\	      
   Si$^+$			&	8.0(-9) $^\textrm{d}$		\\	
   Fe$^+$			&	3.0(-9) $^\textrm{d}$		\\	
   Na$^+$			&	2.0(-9) $^\textrm{d}$		\\	
   Mg$^+$			&	7.0(-9) $^\textrm{d}$		\\	   
   P$^+$			&	2.0(-10) $^\textrm{d}$	\\	      
   Cl$^+$			&	1.0(-9) $^\textrm{d}$		\\	        
   \hline
   \end{tabular}
   \end{center} 
   \medskip
   $^\dagger$ a(b) = a$\times10^{\textrm{b}}$\\
   $^\textrm{a}$ See discussion in \citet{Wakelam08}\\
   $^\textrm{b}$ From \citet{Jenkins09}\\
   $^\textrm{c}$ See discussion in \citet{Hincelin11}\\
   $^\textrm{d}$ Low metal abundance from \citet{Graedel82}   
\end{table}

\subsection{Effect of the Eley-Rideal and complex induced reaction mechanisms}

\subsubsection{Formation of Complex Organic Molecules}
\begin{figure*} 
   \includegraphics[width=84mm,trim = 2cm 3cm 2cm 2cm, clip, width= 6cm, angle=270]{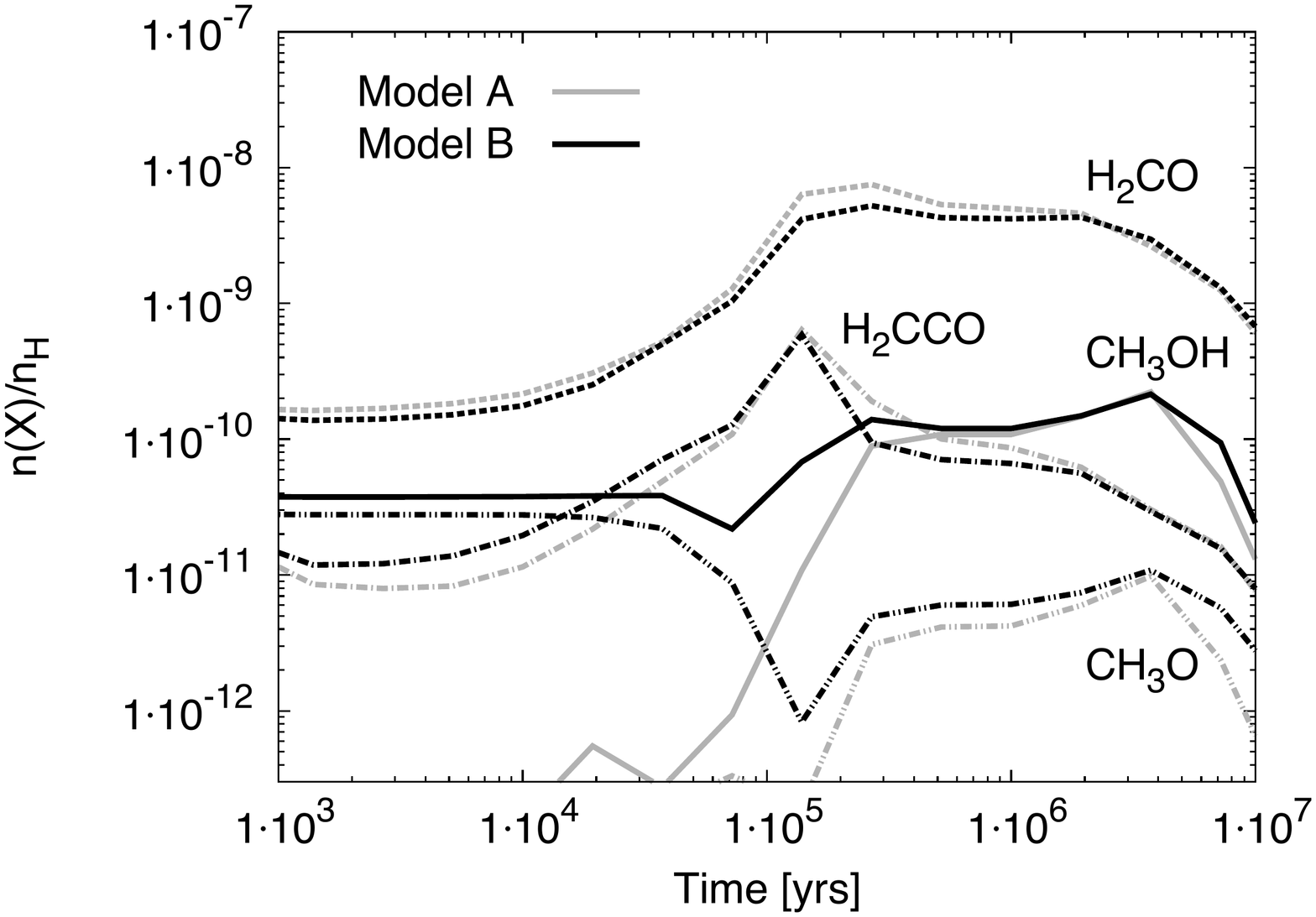}
   \includegraphics[width=84mm,trim = 2cm 3cm 2cm 2cm, clip, width= 6cm, angle=270]{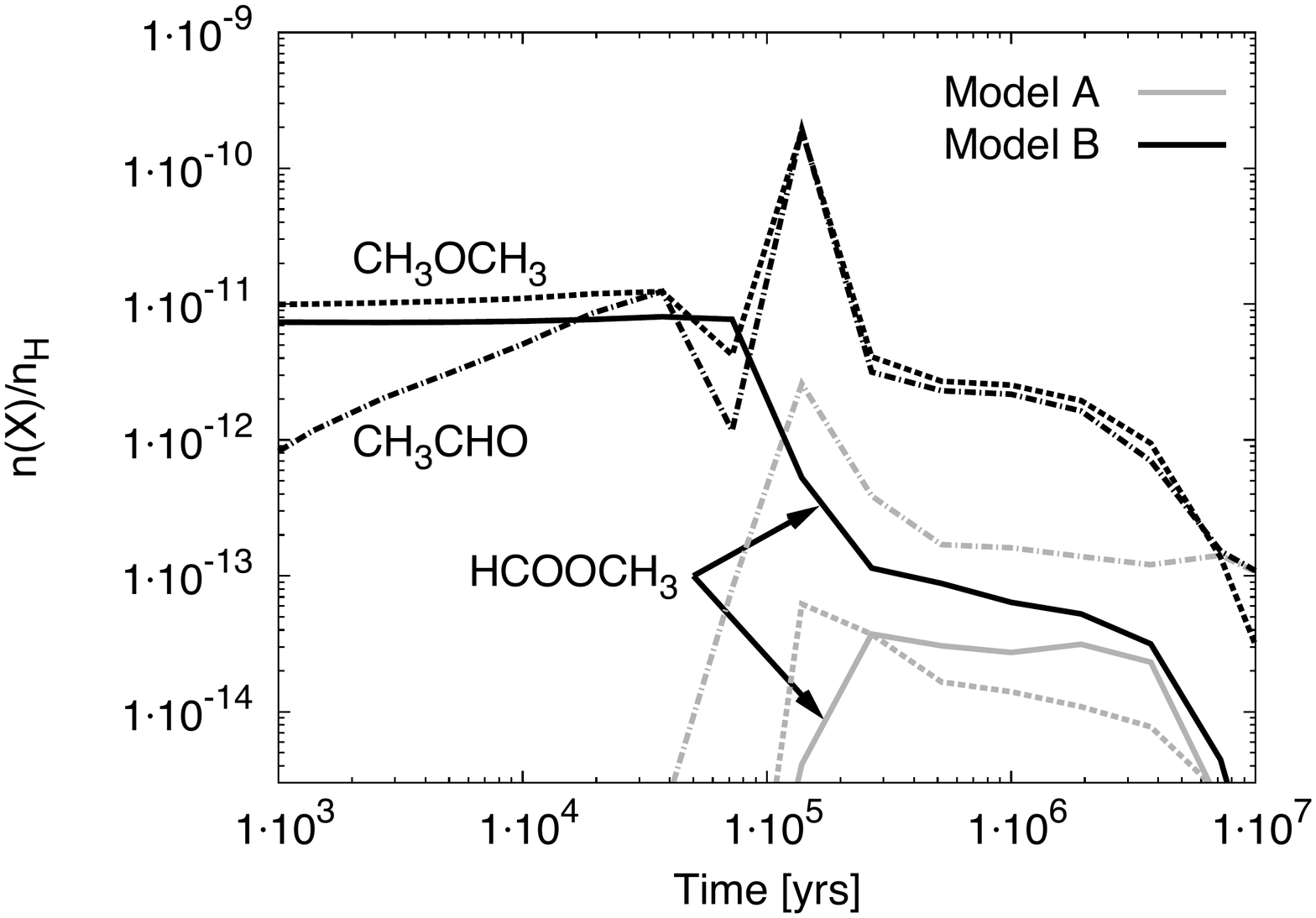}
   \caption{Abundance (with respect to \nh) of selected species in the gas-phase for Model A and B as a function of time.} 
   \label{comp_ab_modA_modB_gas}
\end{figure*}

\begin{figure*} 
   \includegraphics[width=84mm,trim = 2cm 3cm 2cm 2cm, clip, width= 6cm, angle=270]{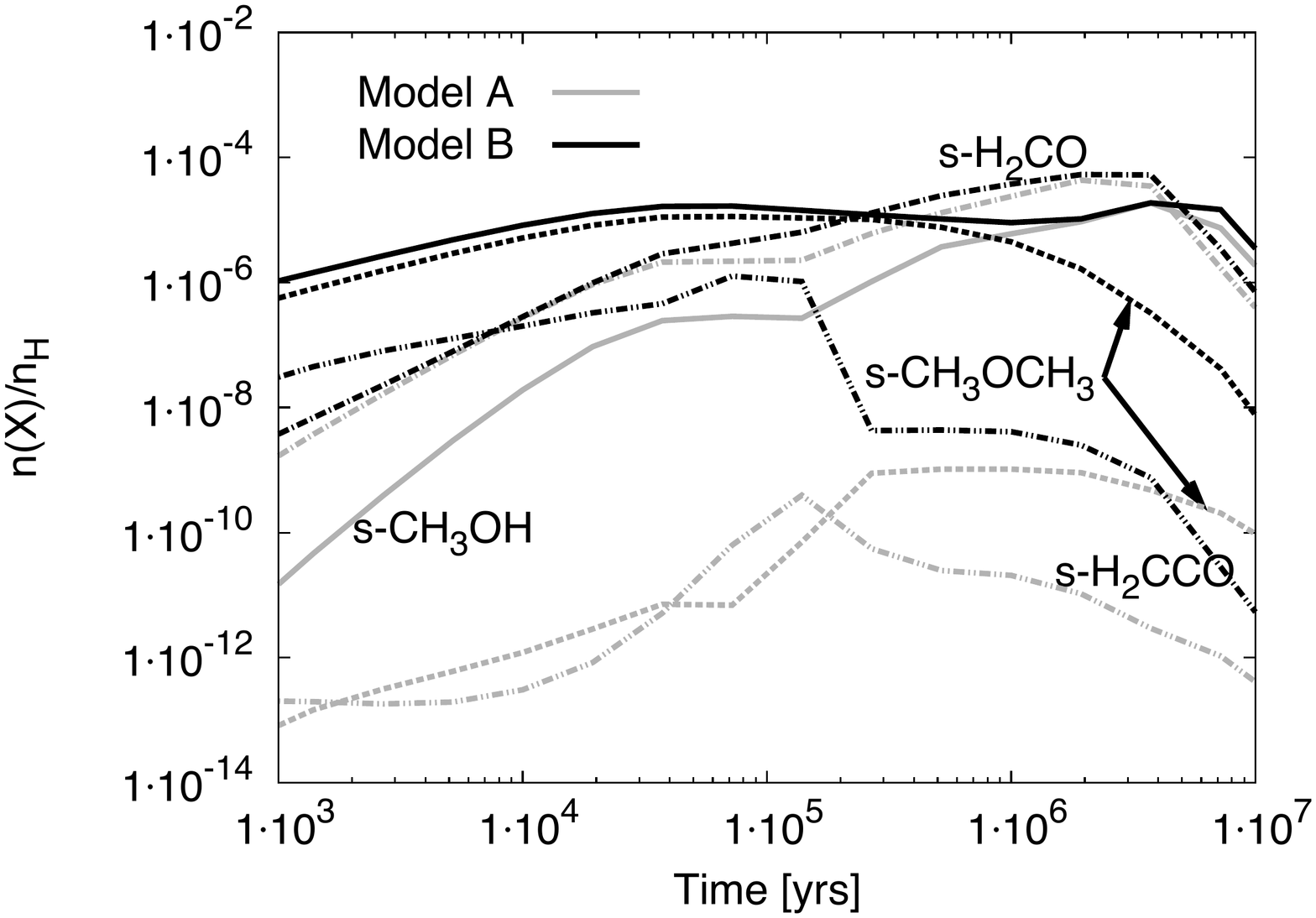}
   \includegraphics[width=84mm,trim = 2cm 3cm 2cm 2cm, clip, width= 6cm, angle=270]{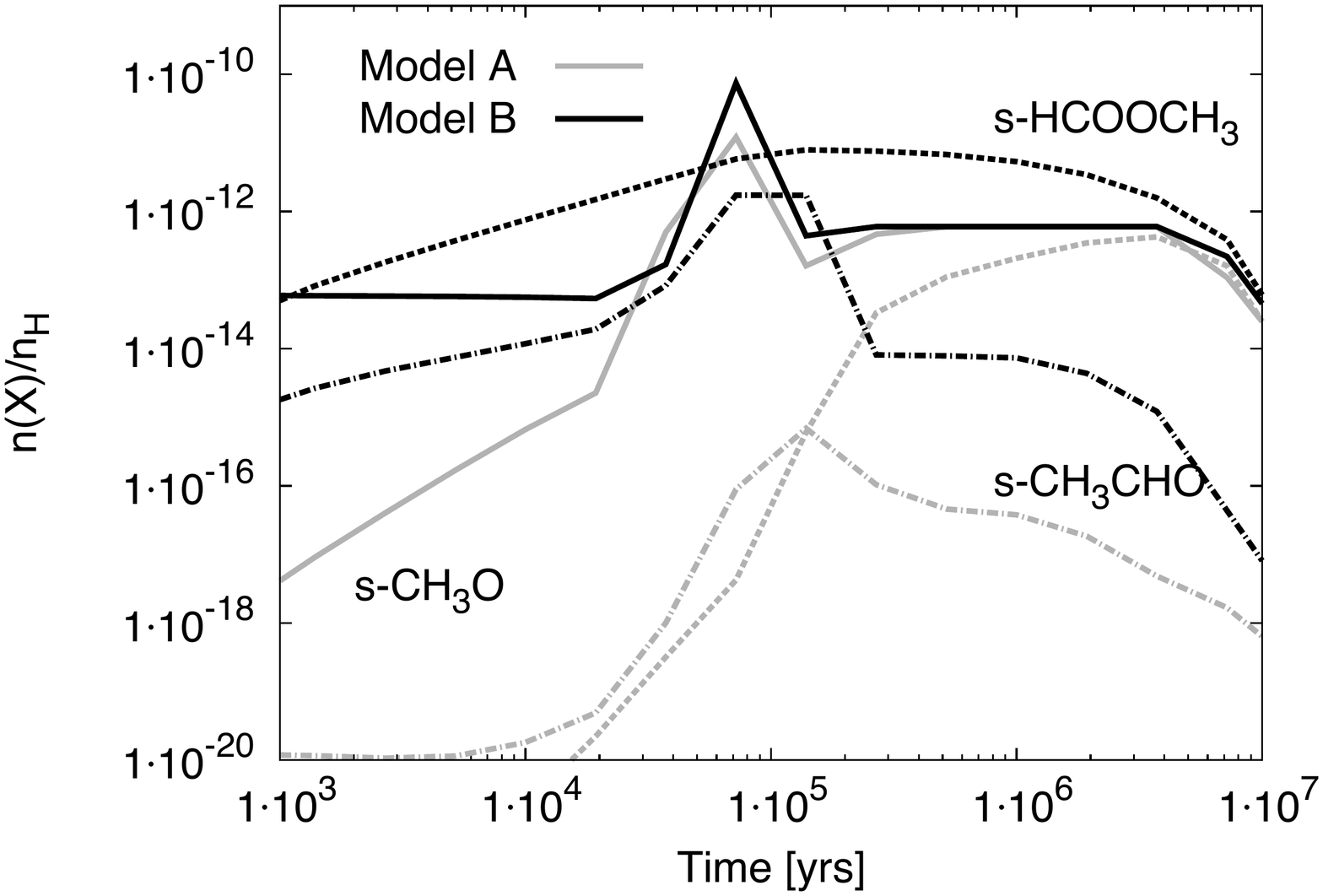}
   \caption{Abundance (with respect to \nh) of selected species on the grain surfaces for Model A and B as a function of time.} 
   \label{comp_ab_modA_modB_grain}
\end{figure*}

Our calculations suggest that the Eley-Rideal and complex induced reaction mechanisms play an important role in the formation of some complex organic molecules.  Figs. \ref{comp_ab_modA_modB_gas} and \ref{comp_ab_modA_modB_grain} show the abundances of a selection of these species (\chhhoh, \hhco, H$_2$CCO, HCOOCH$_3$, CH$_3$OCH$_3$, CH$_3$CHO and CH$_3$O) as a function of time in the gas-phase and at the surface of the grains, for Models A (nominal model) and B (model with the new mechanisms). The \hhco~ abundance is not changed in the gas-phase or on the grains. The H$_2$CCO abundance on the surface is strongly increased whereas in the gas it is unchanged. The gas phase abundances of H$_2$CO and H$_2$CCO do not change much because these species are efficiently produced in the gas phase 
following the forming process described in \citet{Vasyunin13}, i.e. O + CH$_3$ for H$_2$CO, O + C$_2$H$_3$ and the dissociative recombination of CH$_3$CO$^+$ for H$_2$CCO. For the other selected species, Model B shows a strong enhancement both in the gas-phase and at the surface of the grains.

\begin{figure} 
   \includegraphics[width=84mm,trim = 2cm 3cm 2cm 2cm, clip, width= 6cm, angle=270]{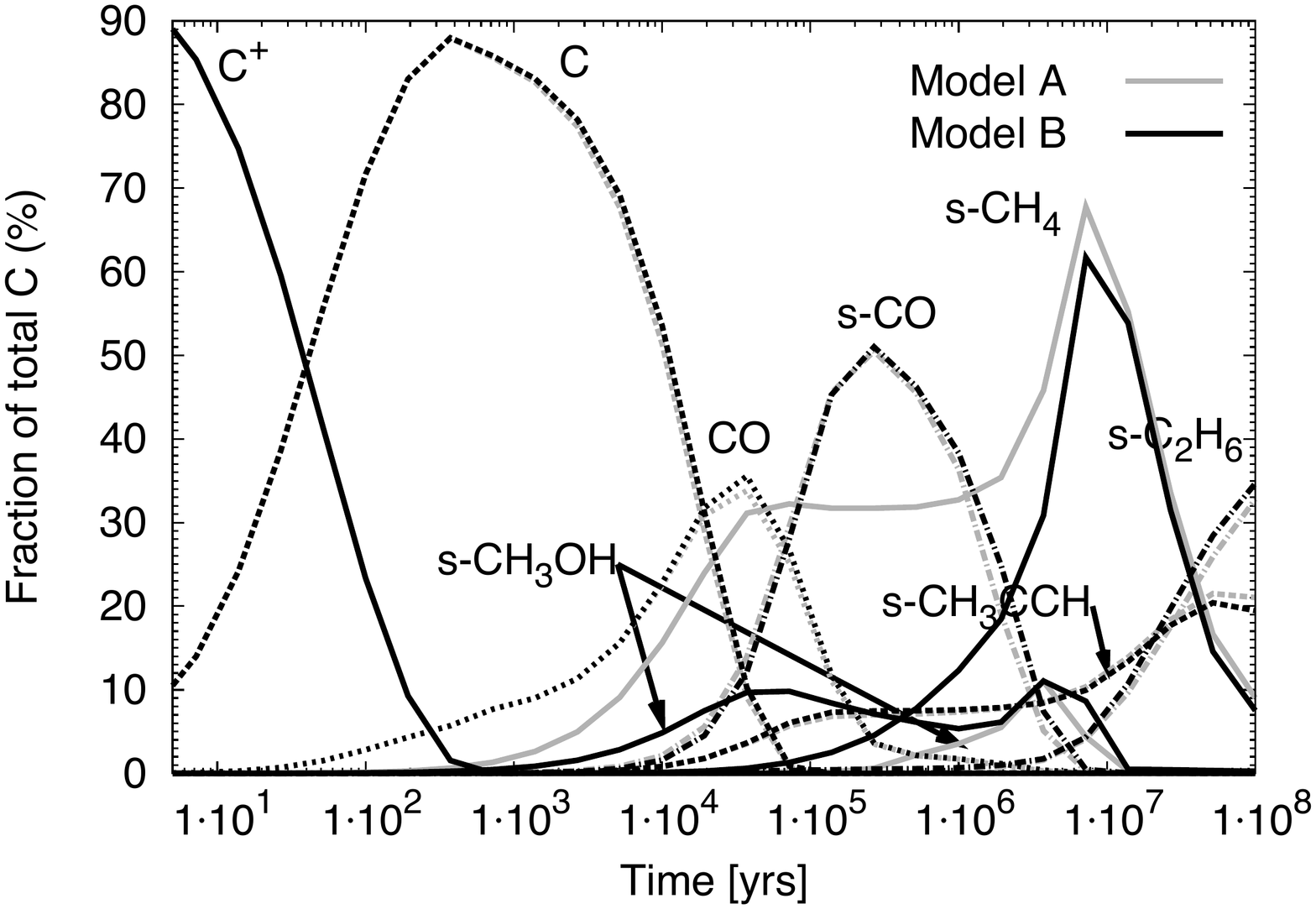}
   \caption{Main carriers of carbon as a function of time for Model A and B.} 
   \label{carbon}
\end{figure}

Before $10^4$ yr, carbon is mostly in atomic form (see Fig.~\ref{carbon}). Thus the formation of C complexes with the main constituants of ices at these times, i.e. H$_2$O and CO, is very efficient and produces larger molecules more quickly than in Model A. The larger abundance of these species in the gas-phase then reflects the larger surface abundances.

In Model B, the formation of methanol starts on the surface by the formation of the s-C...H$_2$O complex through the adsorption of gas-phase atomic carbon on the water ice covered grain. The hydrogenation of this complex leads to a reorganization of the molecule to form s-CH$_2$OH and s-CH$_3$O. Hydrogenation of these molecules produce methanol:
\begin{equation}
  \textrm{C} + \textrm{s-H}_2\textrm{O} \rightarrow \textrm{s-C...H}_2\textrm{O} \xrightarrow{\textrm{s-H}} \textrm{s-CH}_2\textrm{OH}/\textrm{s-CH}_3\textrm{O} \xrightarrow{\textrm{s-H}} \textrm{s-CH}_3\textrm{OH}
\end{equation}
Similarly, the CH$_3$OCH$_3$ molecule is formed on the surface through the following path:
\begin{equation}\label{CH3OCH3}
  \textrm{C} + \textrm{s-CH}_3\textrm{OH} \rightarrow \textrm{s-C...CH}_3\textrm{OH} \xrightarrow{\textrm{s-H}} \textrm{s-CH}_3\textrm{OCH}_2 \xrightarrow{\textrm{s-H}} \textrm{s-CH}_3\textrm{OCH}_3
\end{equation}
At early times, the s-CH$_3$OCH$_3$ abundance is increased because s-CH$_3$OH is more abundant and because the direct formation of s-C...CH$_3$OH by van der Waals complexes is more efficient than by a diffusive Langmuir-Hinshelwood mechanism.

The s-CH$_3$CHO molecule is formed by:
\begin{equation}\label{CH3CHO}
\begin{split}
  \textrm{C} + \textrm{s-CO} \rightarrow \textrm{s-CCO}  \xrightarrow{\textrm{s-H}} \textrm{s-HC}_2\textrm{O} \xrightarrow{\textrm{s-H}} \textrm{s-H}_2\textrm{CCO} \xrightarrow{\textrm{s-H}} \textrm{s-CH}_3\textrm{CO} \\ \xrightarrow{\textrm{s-H}} \textrm{s-CH}_3\textrm{CHO}.
\end{split}
 \end{equation}
 Here again, the s-CH$_3$CHO abundance is much larger compared to Model A because the precursor s-CCO is much more efficiently formed. 
 
Although we did not change the formation of HCOOCH$_3$ on the surface, its abundance is strongly increased. In fact, during the formation of the COMs on the surface, a small fraction of the products (including intermediate radicals) desorb into the gas-phase due to reactive desorption. These radicals can then undergo gas-phase reactions and lead to the formation of more complex molecules \citep[see also][]{Vasyunin13}. Before $10^4$ yr, HCOOCH$_3$ is formed by the reaction between atomic oxygen and CH$_3$OCH$_2$ (from reaction path \eqref{CH3OCH3}). After $10^4$~yr, it is the dissociative recombination of H$_5$C$_2$O$_2^+$, itself formed from reaction between formaldehyde and CH$_3$OH$_2^+$, which produces HCOOCH$_3$.

After a few $10^5$~yr, other gas-phase reactions contribute to the formation of CH$_3$OCH$_3$ and CH$_3$CHO, such as the dissociative recombination of CH$_3$OCH$_4^+$ and CH$_3$CHOH$^+$ and the neutral-neutral reaction O + C$_2$H$_5$ $\rightarrow$ CH$_3$CHO + H. However, the formation paths \eqref{CH3OCH3} and \eqref{CH3CHO} are always the dominant ones. It is interesting to point out that despite the fact that we have not changed the desorption mechanisms, the increase of the surface abundances are large enough to propagate to the gas-phase abundances. We have only considered a small efficiency for the chemical desorption (see section \ref{gas_grain_code}). Increasing this efficiency, would increase the gas-phase abundances of COMs by the same amount.
Moreover, we have introduced the reactions of carbon atoms with H$_2$CO and CH$_3$CHO (as well as with CH$_3$OCH$_3$ and H$_2$CCO) using \citet{Husain99}. These new reactions are rapid and decrease the abundance of these species as long as gas phase C is abundant (few $10^5$~yr).

\subsubsection{Effect on the major ice compounds}
\begin{figure} 
   \includegraphics[width=84mm,trim = 2cm 3cm 2cm 2cm, clip, width= 6cm, angle=270]{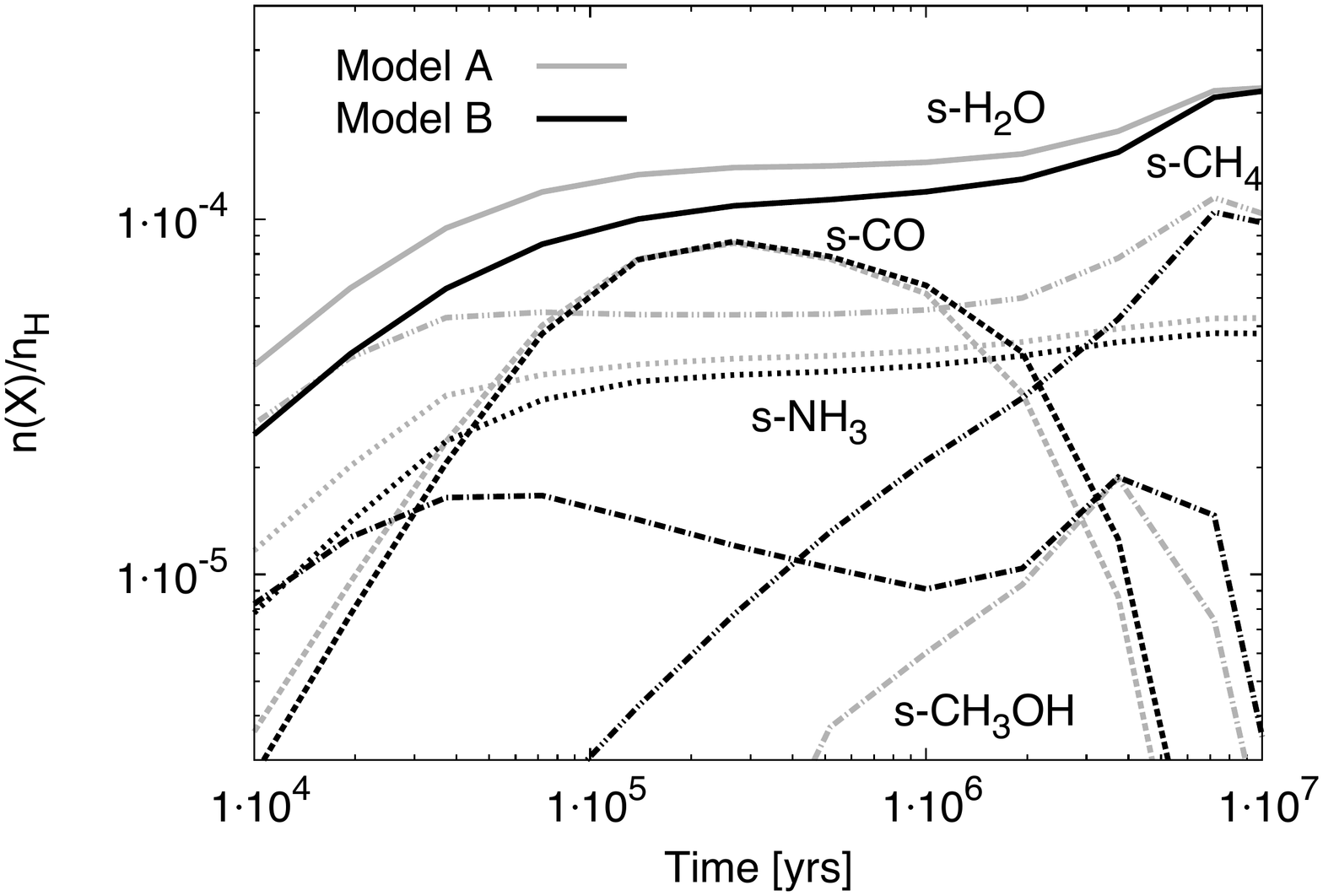}
   \caption{Major ice compounds abundance as a function of time for Models A and B.} 
   \label{ice_composition}
\end{figure}
Figure \ref{ice_composition} shows the abundances of the major ice compounds as a function of time for Models A and B. As expected the major ice component is water, with a fractional abundance $\ge 10^{-4}$ after $10^6$ yr, which corresponds to approximately 100 monolayers deposited on each grain. CO becomes the second major component between $8\times10^4$ and $1\times10^6$ yr due to a massive freeze-out from the gas at these times. The modification of the chemical networks in Model B has a moderate impact on the ice composition compared to our nominal version except for s-CH$_4$ and s-CH$_3$OH.

The s-CH$_3$OH abundance is enhanced by several orders of magnitude until $2\times10^6$ yr, where both models give similar abundances. This difference is due to the efficient formation mechanism of s-CH$_3$OH through successive hydrogenation of s-C...H$_2$O as described before. In contrast s-CH$_4$ is strongly decreased until $5\times10^6$ years. This trend can be explained by the difference in the s-CH abundance, s-CH$_4$ being produced by the hydrogenation of s-CH in both models. In the new network, s-CH is formed by one of the product channels of the reactions between atomic hydrogen s-H and complexes with atomic carbon (s-C...H$_2$O, s-C...NH$_3$, s-C...CO$_2$ etc, see Table \ref{grain_chemistry} of Appendix \ref{network}). The branching ratio producing s-CH for such channels have been set to 1\% (see section \ref{chem}). The consequence is that most of the carbon is then transferred into more complex species and the abundance of s-CH is much smaller compared to Model A, in which s-CH comes from the sticking of gas-phase CH. As a test, we have increased this branching ratio and the result is that the s-CH$_4$ increases until it produces the same amount of s-CH$_4$ as in Model A if the branching ratio is set to 100\%.

\subsubsection{The CO$_2$ ice case}

Carbon dioxide is one of the most abundant species observed in interstellar ices ranging into the tens of percent with respect to the water ice abundance \citep[see][]{Oberg11}. Chemical models have difficulties to reproduce such large abundances of a species that need heavy species to diffuse on the cold interstellar grains before forming CO$_2$. Indeed, three processes are typically cited for the formation of CO$_2$ on grain surfaces \citep[see][]{Garrod11}:
\begin{equation}\label{CO2_1}
   \textrm{s-HCO} + \textrm{s-O} \rightarrow \textrm{s-CO}_2 + \textrm{H},
\end{equation}
\begin{equation}\label{CO2_2}
\textrm{s-CO} + \textrm{s-OH} \rightarrow \textrm{s-CO}_2 + \textrm{H},
\end{equation}
\begin{equation}\label{CO2_3}
\textrm{s-CO} +\textrm{s-O} \rightarrow \textrm{s-CO}_2.
\end{equation}

Reaction \eqref{CO2_1} is barrierless but s-HCO is quickly hydrogenated to form s-H$_2$CO. Reaction \eqref{CO2_2} is typically assigned a small activation energy barrier but s-CO and s-OH need to be mobile on the grain surface, which is not the case at 10K considering diffusion by thermal hopping only. Reaction \eqref{CO2_3} is thought to require a more substantial activation energy (i.e. 1000K) but it happens to be the main formation path of CO$_2$ ice in Model A. Following \citet{Garrod11} a new path is introduced in Model B:
\begin{equation}
\textrm{O} + \textrm{s-CO} \rightarrow \textrm{s-O...CO} \xrightarrow{\textrm{s-H}} \textrm{s-OH...CO}  \rightarrow \textrm{s-CO}_2 + \textrm{s-H}
\end{equation}

\begin{figure} 
   \includegraphics[width=84mm,trim = 2cm 3cm 2cm 2cm, clip, width= 6cm, angle=270]{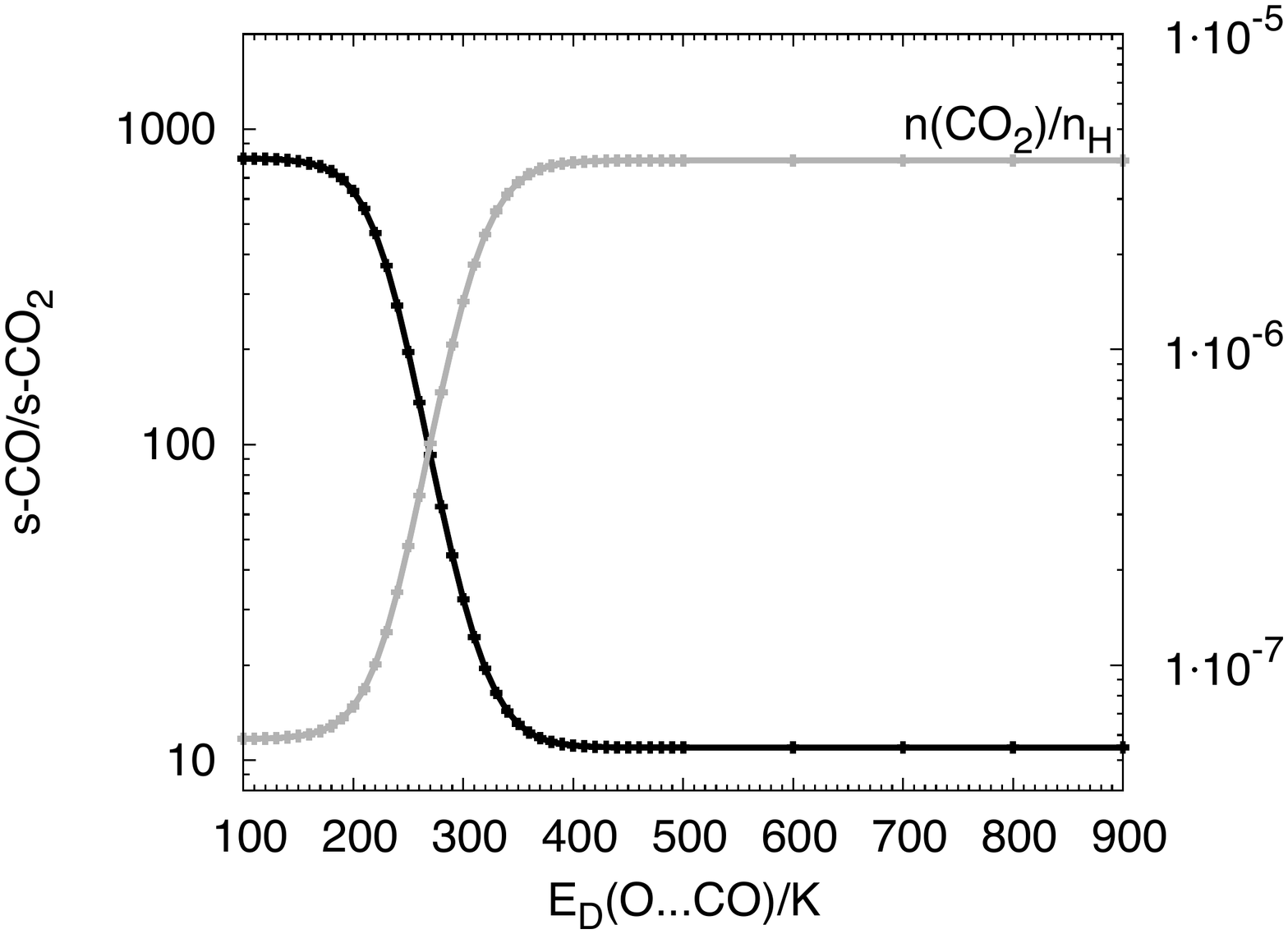}
   \caption{Computed s-CO/s-CO$_2$ ratio (at $10^6$ yr) and CO$_2$ abundance on the grain surface as a function of the binding energy of s-O...CO(K).} 
   \label{ED_O-CO}
\end{figure}

The efficiency of this formation mechanism depends on the binding energy of the s-O...CO complex. In Model B, we have assumed a value of 258~K based on \citet{Goumans10}. This value is calculated at the MPWB1K and CCSD(T)/CBS levels, which are thought to describe weak interactions well. However the calculations concern free CO and the binding value can be notably different for CO adsorbed on ice. To illustrate the effect of this assumption, we have varied this binding energy over the range 100-900~K. Fig.~\ref{ED_O-CO} presents the computed s-CO/s-CO$_2$ ratio as a function of E$_D$(s-O...CO) for t=$10^6$ yr. The efficiency of the conversion of s-CO to s-CO$_2$ strongly decreases with the binding energy of the s-O...CO complex and reaches a plateau when  E$_D$(s-O...CO) is larger than 350~K. When the binding energy of s-O...CO is smaller than $\simeq$200K, the life time of the s-O...CO complex on the grain surface is too short due to thermal evaporation and the complex is destroyed before reacting (i.e. for the set of parameters chosen here). When E$_D$$\simeq$350K, the oxygen is strongly bound to the CO and the life time of the O...CO complex on the grain surface is sufficiently long to form efficiently s-CO$_2$ by the introduced pathway (s-CO/s-CO$_2$$\simeq$10). In term of grain coverage, when E$_D$(s-O...CO)$\gtrsim$400K $\sim$ 2 monolayer of CO$_2$ ice are built on the grain surface.

\subsection{Sensitivity to the model physical parameters}
In this section, we study the sensitivity of our new model to the temperature, the cosmic ray ionization rate and the density. 

\subsubsection{Sensitivity to the temperature and the cosmic ray ionization rate}

We have run Models A and B with a temperature of T=8 and 12 K (smaller and larger than in our nominal one) and for a cosmic ionization rate $\zeta_{\textrm{H}_2}=5.0\times10^{-18}$ and $5.0\times10^{-17}$~s$^{-1}$ (again smaller and larger than the nominal model). For each model, all the other input parameters are kept constant and equal to the standard ones defined in Tables \ref{cold_dense_cloud_model} and \ref{initial_abundances}.

With a lower temperature (8K instead of 10~K), the abundances of methanol and the methoxy radical (CH$_3$O) are lowered by a factor of $\sim10$ between t=$10^5$ and $10^6$ yr in the gas phase. At this temperature, the diffusion of atomic hydrogen is much less efficient than in the T=10 K case. This result is independent of the new mechanisms that we have introduced. At a temperature of 12~K instead of 10~K, the species abundances are not significantly affected, very likely because the increase of the surface reactivity is counterbalanced by the increase in the evaporation of atomic hydrogen from the surface.

Varying the cosmic ionization rate has a significant impact on the computed abundances of all the studied species. In particular, an increase in this rate increases the formation of radicals at the surface of the grains by cosmic-ray induced UV photodissociations and increases the evaporation of surface species through the stochastic heating of grains. In our case, this promotes the surface radical coverage and helps to desorb molecules into the gas-phase. On average, the gas phase abundance of the studied species is enhanced by a factor of $\sim$2-10 between $10^5$ and $10^6$ yrs when $\zeta_{\textrm{H}_2}$ increases from $5.0\times10^{-18}$ to $5.0\times10^{-17}$ s$^{-1}$. This result was however also obtained with the older chemical network (Model A).

\subsubsection{Sensitivity to the density}

\begin{figure*} 
   \includegraphics[width=84mm,trim = 2cm 3cm 2cm 2cm, clip, width= 6cm, angle=270]{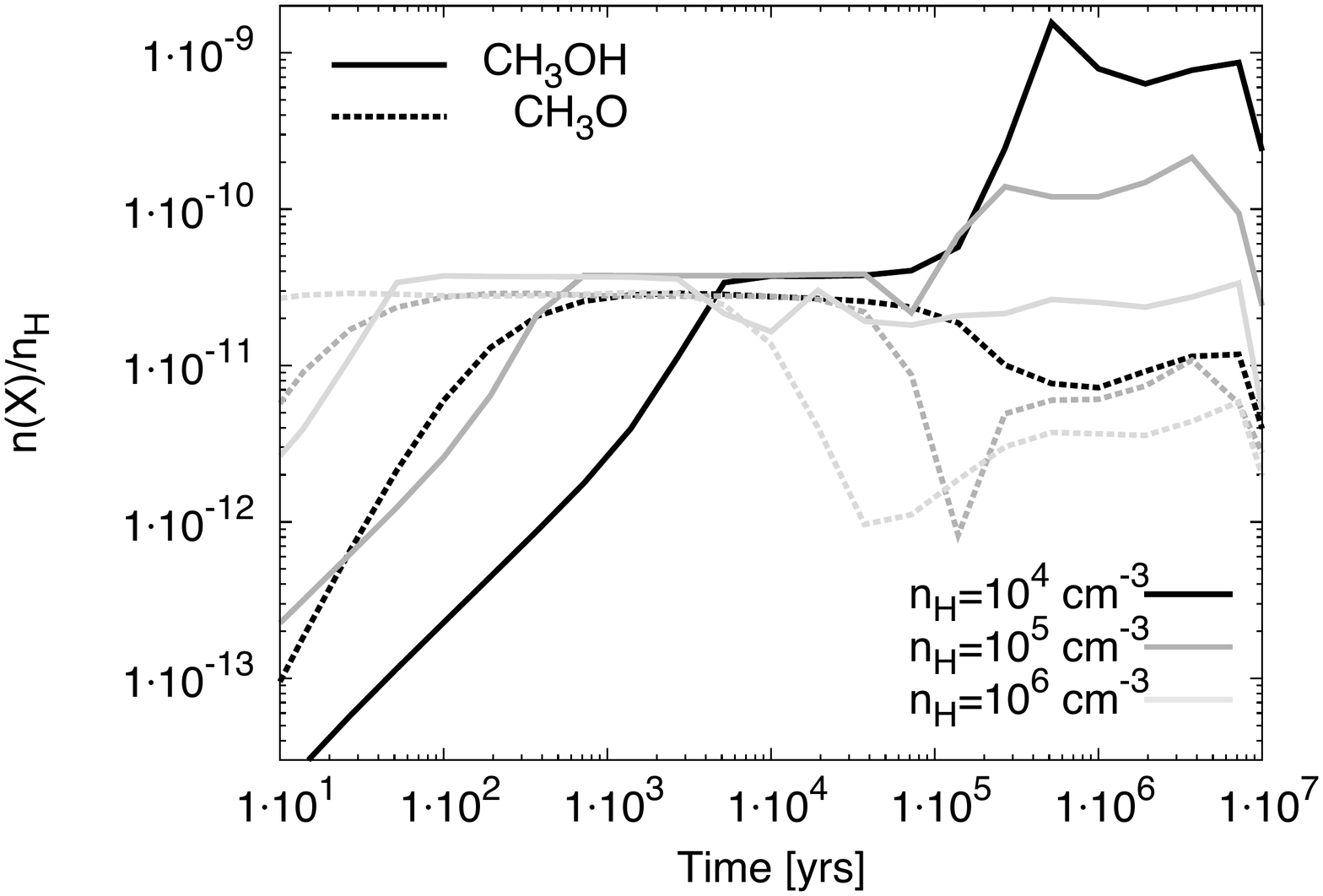}
   \includegraphics[width=84mm,trim = 2cm 3cm 2cm 2cm, clip, width= 6cm, angle=270]{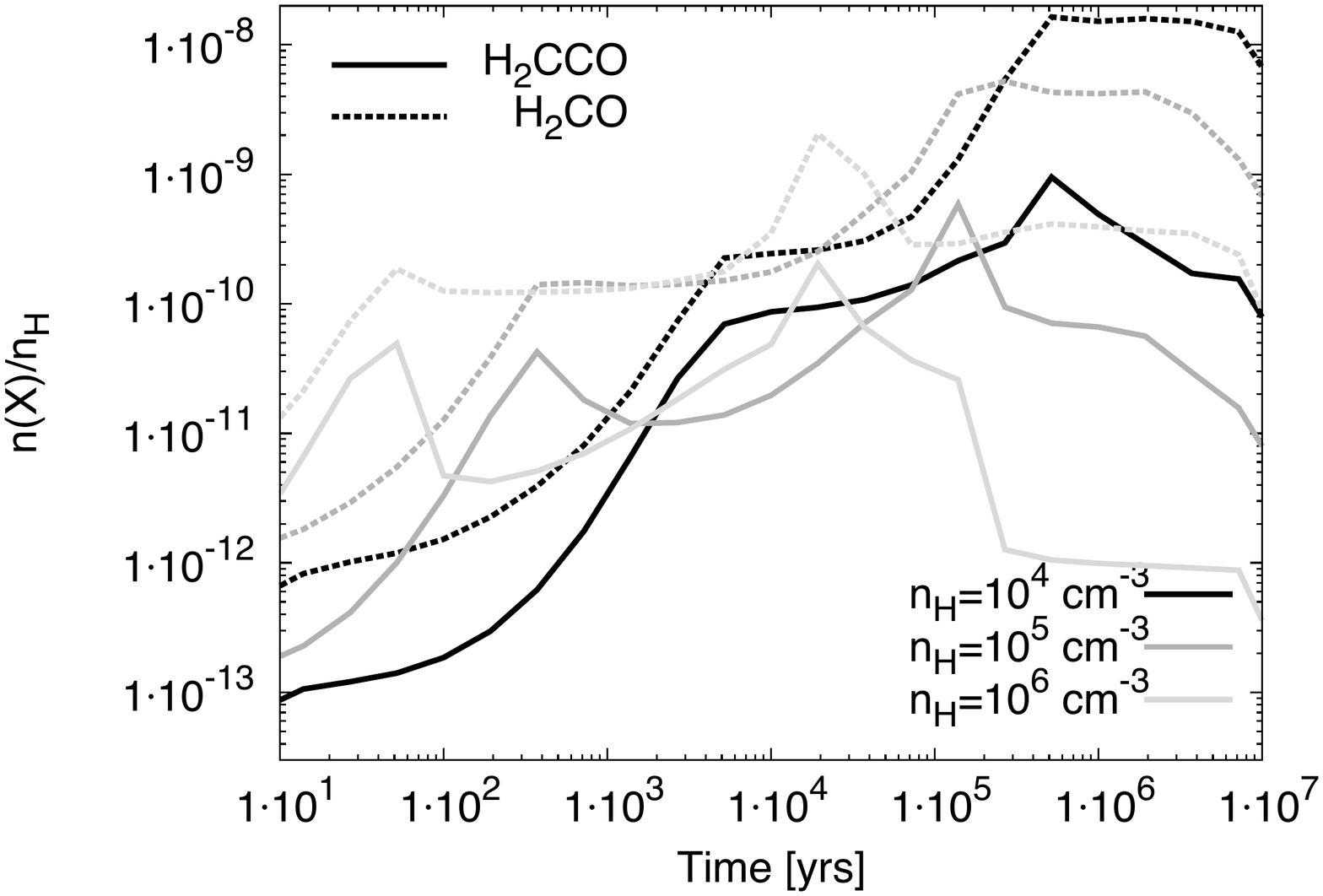}
   \caption{Abundance (with respect to \nh) as a function of time of selected gas-phase species for Model B with \nh=$1.0\times10^4$, $1.0\times10^5$ and $1.0\times10^6$ \cmt.} 
   \label{density}
\end{figure*}

We ran Models A and B for densities \nh=$1.0\times10^4$ and $1.0\times10^6$ \cmt, in addition to the standard density of \nh=$1.0\times10^5$~\cmt. All the other input parameters were kept constant and equal to the standard ones defined in Tables \ref{cold_dense_cloud_model} and \ref{initial_abundances}. Fig.~\ref{density} shows the abundances of CH$_3$OH, CH$_3$O, H$_2$CCO and H$_2$CO in the gas-phase as a function of time for Model B and the three densities. In the gas-phase, as well on the grain surface, the abundance of these species rise earlier. This is due to a larger depletion of species onto the grains and thus a greater production of COMs. However, in the gas-phase, the abundance reach a plateau (similar for all densities) while the abundance on the grain surface is still increasing. This is due to the increase of the destruction by reactions with gas-phase atomic carbon (which is the main destruction path until carbon is in atomic form) with the density. This effect is only seen in Model B since with the introduced mechanisms the formation of COMs at early times is completely controlled by the abundance of gas-phase atomic carbon. At longer times ($\le 10^5$~yr), the abundances are larger when the density is smaller simply because the species tend to stay on the grains at larger densities.

\subsection{Sensitivity to the reaction s-CO + s-H}
The efficiency of the formation of methanol on the surface by hydrogenation of CO ice is highly debated and there is still no real consensus. Initially \citet{Tielens82} and \citet{Tielens87} proposed the following hydrogenation scheme from CO to CH$_3$OH \citep[see][]{Tielens97}: 
\[
  \textrm{s-CO}  \xrightarrow{\textrm{s-H}} \textrm{s-HCO} \xrightarrow{\textrm{s-H}} \textrm{s-H}_2\textrm{CO}  \xrightarrow{\textrm{s-H}} \textrm{s-CH}_3\textrm{O} \xrightarrow{\textrm{s-H}} \textrm{s-CH}_3\textrm{OH},
\]
Following this idea, several experimental studies of H-atom bombardment of CO ice were performed by different groups leading to conflicting results \citep{Hiraoka02,Watanabe02}. Both, H + CO and H + H$_2$CO reactions, show notable barriers for reaction in the gas-phase and this is very likely to be the case on grain surfaces. However, on grain surfaces, H is expected to undergo efficient quantum tunneling. \citet{Fuchs09} found that the discrepancy between these two groups was caused mainly by a difference in the H atom flux setting. They also found that the s-CO + s-H and s-H$_2$CO + s-H hydrogenation reactions proceed by a tunneling process with an energy barrier of $390\pm40$K and $415\pm40$K respectively (experience performed with a 12K interstellar ice analog).

\begin{figure*} 
   \includegraphics[width=84mm,trim = 2cm 3cm 2cm 2cm, clip, width= 6cm, angle=270]{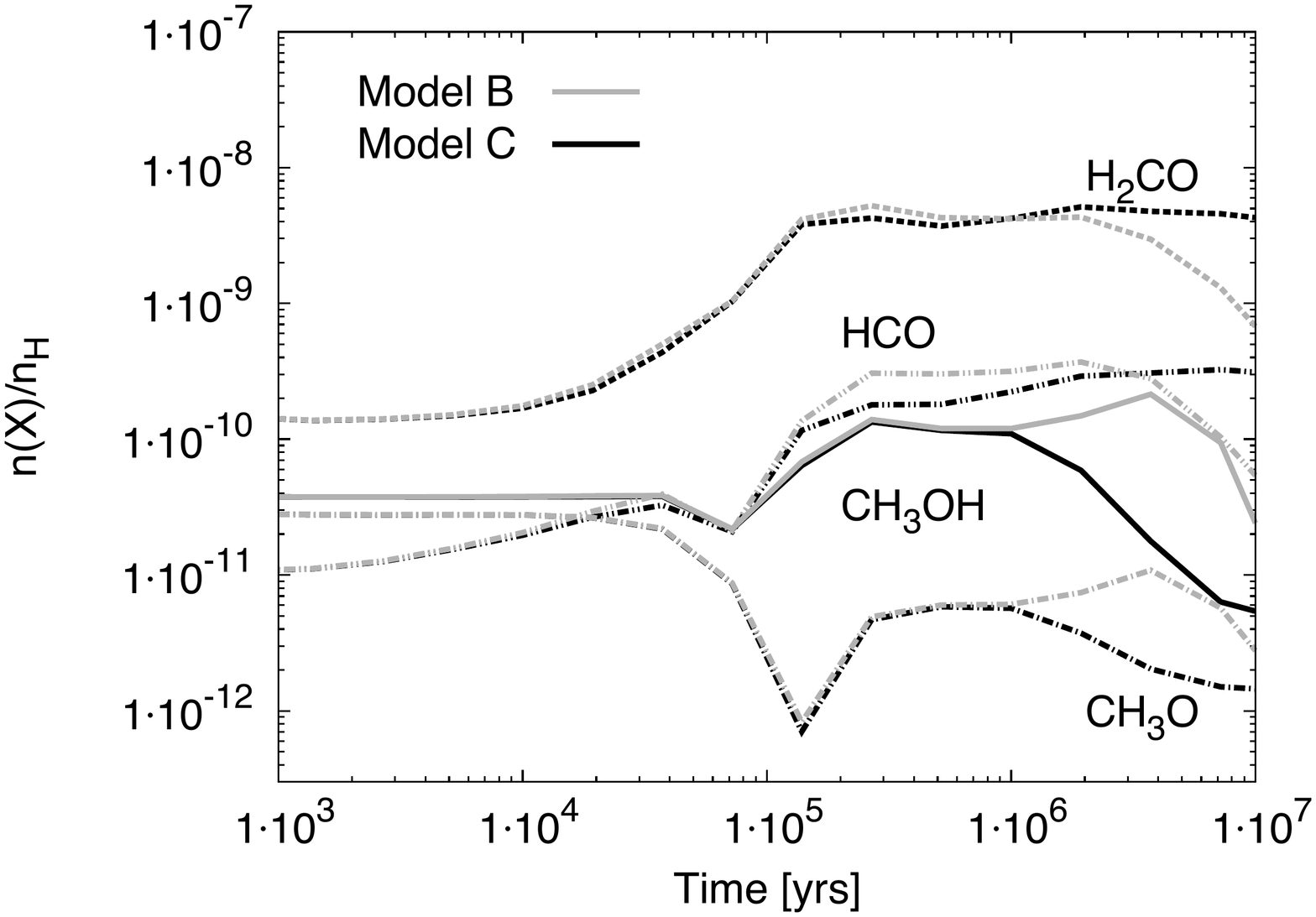}
   \includegraphics[width=84mm,trim = 2cm 3cm 2cm 2cm, clip, width= 6cm, angle=270]{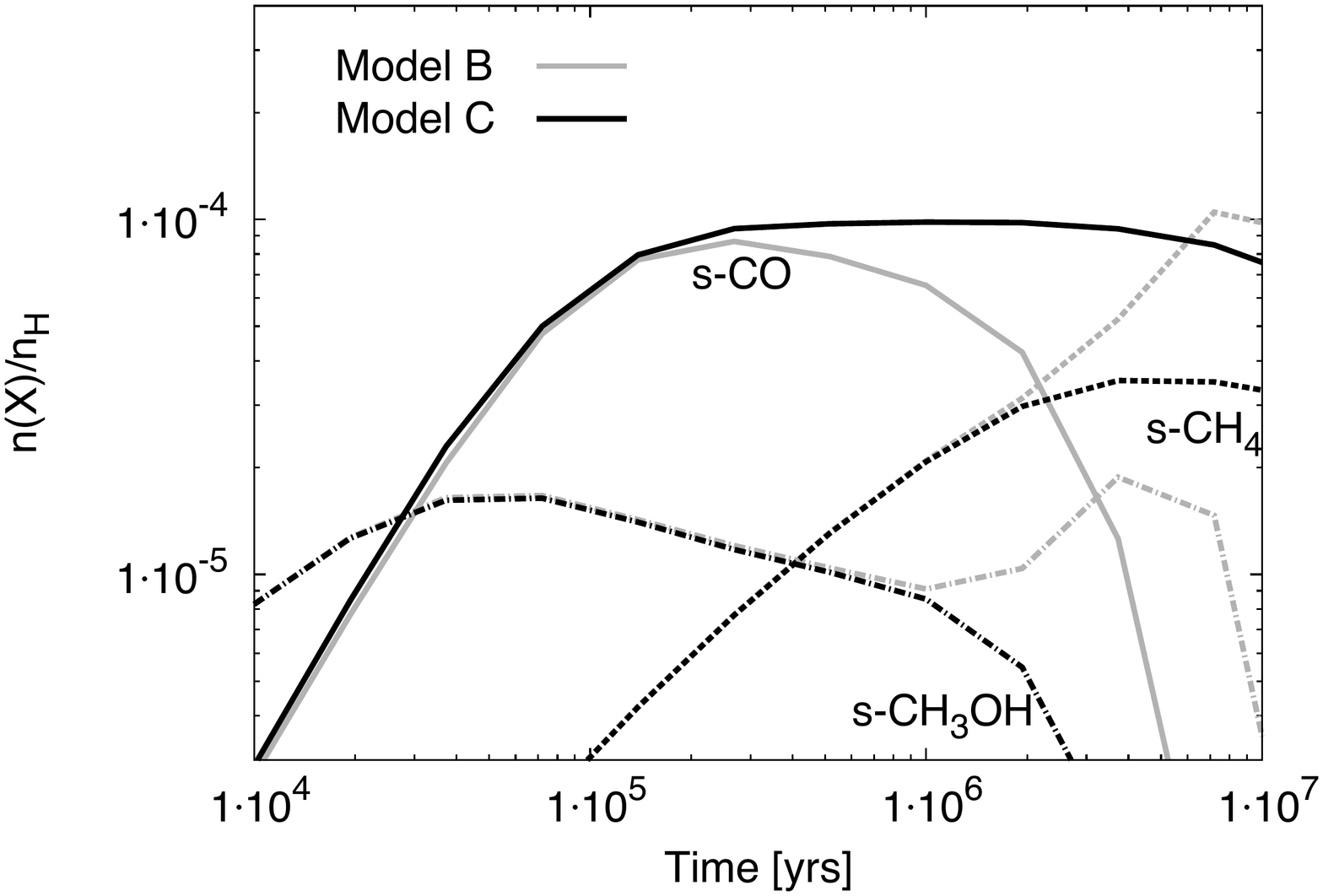}
   \caption{Abundance (with respect to \nh) as a function of time of selected gas-phase and grain surface species for Model B and C. Models ran using standard parameters defined in Tables \ref{cold_dense_cloud_model} and \ref{initial_abundances}.} 
   \label{test_H_CO}
\end{figure*}

In order to test this methanol formation path against our new mechanism, we ran an additional model in which we switched off the reaction s-CO + s-H in our Model B (i.e. which is equivalent to making the assumption that the thickness of the barrier cannot be overcome by tunneling). Fig. \ref{test_H_CO} shows the computed abundances of s-CO, s-HCO and s-CH$_3$OH for Model B and the additional model where the Eley-rideal  and complex induced reactions mechanisms were allowed and the s-H + s-CO reaction turned off (Model C). Before $10^5$yr, all computed abundances are similar in both models. After $10^5$ yr, removing the hydrogenation of s-CO (Model C) increases strongly the abundance of s-CO.  The abundance of CH$_3$OH in the gas-phase and on the grain surfaces is not too affected until $\sim10^6$~yr. As a conclusion, the new formation path of s-CH$_3$OH that we have introduced is found to be the dominant one until $\sim10^6$~yr and is able to produce a large amount of methanol on grain surfaces and in the gas-phase.

\section{Comparison with observations in cold dense cores}
To compare the modeling results with observations, we used the method described in \citet{Loison14}. We have computed at each time step the mean distance of disagreement using the following formula: 
\begin{equation} \label{eq:distance}
   D(t) = \frac{1}{N_{obs}} \sum_i \vert~log[n(X)_i(t)] - log[n(X)_{obs}]~\vert,
\end{equation}
where $n(X)_i(t)$ is the calculated abundance of species i and $n(X)_{obs}$ the observed one. $N_{obs}$ refers to the total number of observed species. With this method, the smaller the D value, the better the agreement between modeled and observed abundances.

In order to validate our introduced mechanism, we compared our modeling results with observations of the TMC-1 and L134N dark clouds for which many observational constraints are available (approximately 50 molecules for L134N and 60 for TMC-1 \citep[see][for a review]{Agundez13}). We ran models A and B with a density of n$_\textrm{H}=2\times10^4$ \cmt, all the other parameters being equal to those of Tables \ref{cold_dense_cloud_model} and \ref{initial_abundances}. We found a similar level of agreement between Model A and Model B with a "best age", using equation \eqref{eq:distance}, of $2\times10^6$ yr and $1\times10^6$ yr for for TMC-1 and L134N respectively. At these times, and assuming that the observed abundances were reproduced by the model when the difference between the two was smaller than one order of magnitude, the fraction of reproduced molecule is $\sim60$ and $\sim70\%$ for TMC-1 and L134N respectively. The fact that both Models A and B produce similar agreements is due to the fact that the introduced mechanism does not greatly affect the molecules observed in these two clouds. 

\subsection{COMs in the TMC-1 dark cloud}
Using the data from the 8 GHz to 50 GHz spectral survey of TMC-1 obtained with the 45m Nobeyama radiotelescope \citep{Kaifu04} we determined observational upper limits for CH$_3$OCH$_3$, CH$_3$NH$_2$ and HCOOCH$_3$. For this, the published noise values for each individual line of the survey were interpolated to create a regular noise vector spanning the observed frequency range. We used the WEEDS package \citep{Maret11} to create synthetic LTE spectra with the following common characteristics: a FWHM linewidth of $\Delta_v= 0.5$ km.s$^{-1}$ and an excitation temperature of T$_\textrm{ex}$ =5 K. The column density of each modeled species was varied until at least one of the modeled lines had a peak temperature larger than three times the local noise value. Along with a fixed N(H$_2$)=$10^{22}$ \cmd~this defines the upper limit on the abundance of each species. We find that N(CH$_3$OCH$_3$)/N(H$_2$) $<2\times10^{-9}$, N(CH$_3$NH$_2$)/N(H$_2$) $<5\times10^{-9}$ and N(HCOOCH$_3$)/N(H$_2$) $<1\times10^{-9}$. The calculated abundances of these species at the time of "best age" (i.e. determined in the previous section) with Model B (using a density of n$_\textrm{H}=2\times10^4$) gives n(CH$_3$OCH$_3$)/n(H$_2$) $= 8.30\times10^{-12}$, n(CH$_3$NH$_2$)/n(H$_2$) $= 3.60\times10^{-10}$ and n(HCOOCH$_3$)/n(H$_2$) $= 3.00\times10^{-13}$ which is far lower than the upper limits determined. These upper limits are therefore only poor constraints on the calculated COM abundances in this source.

In the two next sections, we focus on other dense cold sources where COMs have been observed.

\subsection{L1689b and B1-b cold dense cores}
In Table \ref{observed_abundances} we reported the observed COM abundances in B1-b and L1689b from \citet{Bacmann12} and \citet{Cernicharo12}, together with the modeled abundances at the "best age".
The modeled abundances have been obtained with Model A and B for a density of \nh=$1.0\times10^5$ \cmt. The "best age" was obtained by computing the distance of disagreement as previously explained and using the molecules listed in Table \ref{observed_abundances}. We found the "best age" to be $\sim1.0\times10^5$ yrs for L1689b and $\sim5.0\times10^5$ for B1-b. 

At the time of best agreement and except for HCOOCH$_3$, most of the computed abundances differ from observational abundances by less than a factor of $\sim10$ for all species when the Eley-Rideal mechanism and complex induced reactions are included (Model B). In the case of L1689b dense core, the observed abundances of CH$_3$OCH$_3$, CH$_3$CHO and H$_2$CCO are reproduced within a factor of 3 whereas H$_2$CO is reproduced within a factor of 8. In the case of B1-b dense core, CH$_3$O, CH$_3$OCH$_3$ and CH$_3$CHO are very well reproduced (within a factor of 3). H$_2$CCO and H$_2$CO are overproduced by a factor of $\sim$10 and $\sim$20 respectively while CH$_3$OH is underestimated by a factor of $\sim$10. For both clouds, Model B reproduces the observed COM abundances much better. Despite the fact that the HCOOCH$_3$ abundance is larger in Model B than in Model A, we still underestimate the abundance compared to the observations.

\begin{table*}
   \begin{center}
   \caption{Observed and computed fractional abundances of Complex Organic Molecules in L1689b and B1-b dense cores}
   \label{observed_abundances}
   \begin{tabular}{@{}lccccccc}
   \hline
   \hline
		   		&	L1689b$^\dagger$ 	&Model A			&Model B				&&B1-b$^\dagger$ 		&Model A			&Model B\\
  \cline{2-4} \cline{6-8}
   Density (in \cmt)	&					&$1\times10^5$	&$1\times10^5$		&&					&$1\times10^5$	&$1\times10^5$\\
   Time of best fit (in yrs)&					&$2\times10^5$	&$2\times10^5$		&&					&$5\times10^5$	&$5\times10^5$\\
   \hline
   HCOOCH$_3$	&	7.4(-10)			&5.20(-14)		&3.60(-13)			&&	2.0(-11) 			&6.10(-14)		&1.70(-13)\\
   CH$_3$OCH$_3$	&	1.3(-10)			&2.20(-13)		&\textbf{2.35(-10)}		&&	2.0(-11) 			&3.30(-14)		&\textbf{5.50(-12)}\\
   CH$_3$CHO		&	1.7(-10)			&1.89(-12)		&\textbf{1.80(-10)}		&&	1.0(-11) 			&3.40(-13)		&\textbf{4.70(-12)}\\
   H$_2$CCO		&	2.0(-10)			&\textbf{6.80(-10)}	&\textbf{5.50(-10)}		&&	1.3(-11) 			&2.00(-10)		&1.40(-10)\\
   CH$_3$O		&	...				&2.90(-12)		&7.55(-12)			&&	4.7(-12) 			&\textbf{8.30(-12)}	&\textbf{1.20(-11)}\\
   H$_2$CO		&	1.3(-9)			&1.60(-8)			&1.05(-8)				&&	4.0(-10) 			&1.05(-8)			&8.50(-9)\\
   CH$_3$OH		&	...				&1.00(-10)		&3.10(-10)			&&	3.1(-9) 			&2.15(-10)		&2.40(-10)\\
   \hline
   \end{tabular}
   \end{center} 
   \medskip
  Observed and calculated fractional abundances expressed in unit of n(X)/n(H$_2$).  $^\dagger$ Observed fractional abundances are those listed in \citet{Vasyunin13}. Boldface indicate less than a factor of 4 disagreement between model and observations. Here, a(b) stands for a$\times10^{\textrm{b}}$.\\
\end{table*}

\subsection{The prestellar core L1544}
Recently, \citet{Vastel14} reported the detection of a number of COMs in the prestellar core L1544. They estimate, thanks to the methanol lines, that these COM emissions come from the border of the core, at a radius of $\sim 8000$ au, where T $\sim10$ K and n$_{\textrm{H}_2} \sim$ $2.0\times10^4$ \cmt. In this region, \citet{Caselli12} also revealed a large amount of water in the gas phase (i.e. n(\hho)/n$_{\textrm{H}_2}$ $\simeq$ $3.0\times10^{-7}$). The observed and computed abundances for Model A and B (\nh~was set to $4.0\times10^4$ \cmt and all the other input parameters equal to those defined in Tables \ref{cold_dense_cloud_model} and \ref{initial_abundances}) are reported in Table \ref{obs_l1544}. The "best age" was found to be $\sim3.0\times10^5$ yrs in Model A and $\sim2.0\times10^5$ yrs in Model B.

\begin{table}
   \begin{center}
   \caption{Observed and computed fractional abundance of Complex Organic Molecules in L1544 prestellar core}
   \label{obs_l1544}
   \begin{tabular}{@{}lcccc}
   \hline
   \hline
		   		&	L1544$^\dagger$ 		 &Model A			&Model B\\
   \cline{2-4}
   Density (in \cmt)	&						&$4\times10^4$	&$4\times10^4$\\
   Time of best fit (in yrs)&						&$3\times10^5$	&$2\times10^5$\\
   \hline
   HCOOCH$_3$	&	$\leq$1.5(-9)			&2.20(-13)		&4.15(-12)\\
   CH$_3$OCH$_3$	&	$\leq$2.0(-10)			&2.25(-13)		&6.15(-11)\\
   CH$_3$CHO		&	1.0(-10)				&4.15(-12)		&\textbf{6.65(-11)}\\
   HCOOH			&	1.0(-10)				&\textbf{2.25(-10)}	&\textbf{3.75(-10)}\\
   H$_2$CCO		&	1.0(-9)				&\textbf{1.50(-9)}	&\textbf{1.45(-9)}\\
   CH$_3$CCH		&	5.0(-9)				&9.20(-10)		&8.40(-10)\\
   CH$_3$O		&	$\leq$1.5(-10)			&6.90(-12)		&3.40(-12)\\
   CH$_3$OH		&	6.0(-9)				&4.65(-10)		&3.60(-10)\\
   \hline
   H$_2$O			&	3.0(-7)				&\textbf{9.40(-8)}	&\textbf{3.20(-7)}\\
   \hline
   \end{tabular}
   \end{center} 
   \medskip
   Observed and calculated fractional abundances expressed in unit of n(X)/n(H$_2$). $^\dagger$ Fractional abundances are from \citet{Vastel14}. Boldface indicate less than a factor of 4 disagreement between model and observations. Here, a(b) stands for a$\times10^{\textrm{b}}$.\\
\end{table}

In this source also, the computed abundances show a reasonable agreement with the observed ones. In particular the abundances of HCOOH, H$_2$CCO, CH$_3$CHO and H$_2$O are very well reproduced (by less than a factor of 4 between the observed and the calculated abundances). However, we underestimate the abundance of CH$_3$OH and CH$_3$CCH by a factor of $\sim 17$, $\sim 6$ respectively. Our model predicts an abundance of HCOOCH$_3$, CH$_3$OCH$_3$ and CH$_3$O of $4\times10^{-12}$, $6\times10^{-11}$ and $3\times10^{-12}$ respectively.

It is very interesting to note that we do not need the photodesorption process, as suggested by the authors, to reproduce the observed COM and water abundances. In our case, the non-thermal desorption process is driven by a chemical desorption efficiency of 1\%. This value is found to be sufficient to reproduce the observed gas phase abundances thanks to efficient COM and precursor formation on grain surfaces.
\section{Conclusion}

In this paper, we have studied the effect of the Eley-Rideal and complex induced reaction mechanisms of carbon atoms with the main ice components of dust grains on the formation of COMs in cold dense regions. This study was achieved using our gas-grain chemical model {\sevensize NAUTILUS}. We used a moderate value for the efficiency of the chemical desorption (i.e. we considered that $\sim$1\% of the newly formed species dessorb) in agreement with recent experiments made by \citet{Minissale14}. We also studied the effect of oxygen complexation on CO ice on the formation of CO$_2$ ices.

The main conclusions of this modeling study can be summarized as follows:
\begin{enumerate}
\renewcommand{\theenumi}{(\arabic{enumi})}
\item The Eley-Rideal and complex induced reaction mechanisms are found to enhance considerably the abundance of the most complex COMs both in the gas-phase and at the surface of grains. At the time of best agreement, we found that the gas-phase abundances of CH$_3$CHO, CH$_3$OCH$_3$ and HCOOCH$_3$ are enhanced at least by a factor $\sim10$ up to $\sim 100$ when the Eley-Rideal and complex induced reaction mechanisms were activated while the abundances of H$_2$CO, CH$_3$O, CH$_3$OH and H$_2$CCO remains relatively constant. We find that using a moderate value of 1\% for the chemical desorption efficiency, the surface abundances of complex species are large enough to propagate to the gas-phase abundances. The observed and calculated abundances of COMs in the different environments studied here are, in most cases, in good agreement.
\item During this efficient production of COMs on grain surfaces, a small fraction of intermediate radicals are injected into the gas-phase. These intermediate radicals can then undergo gas-phase reactions and contribute to the formation of COMs in the gas-phase. These contributions are found to be efficient at evolved ages (i.e. when the carbon is mostly in molecular form). However, in most cases, the grain surface formation paths are found to be the dominant ones.
\item We introduced the complex s-O...CO for the formation of CO$_2$ ice on grain surfaces. We assumed a binding energy of the s-O...CO complex of 258K based on \citet{Goumans10}. Using this value,  the expected abundance of CO$_2$ ice is not achieved. This is due to the short lifetime of the s-O...CO complex on the grain surfaces (i.e. the complex is destroyed before reacting). We then studied the dependance of the CO/CO$_2$ ice ratios by varying the value of the binding energy of the s-O...CO complex. We find that a value of E$_D\approx400$ K is sufficient to have a significant fraction of CO$_2$ ice on the grain surfaces (i.e. s-CO/s-CO$_2$$\approx 10$).
\item We observe that the main formation route to CH$_3$O is controlled by the grain surface reactions s-C...H$_2$O + s-H and s-H$_2$CO + s-H. These channels are sufficient enough to reproduce the observed abundance of the methoxy radical. In our models the proposed gas-phase formation route CH$_3$OH + OH $\rightarrow$ CH$_3$O + H$_2$O made by \citet{Cernicharo12} and \citet{Vasyunin13} is inefficient compared to those mentioned above.
\item The methanol formation route by successive hydrogenation of the s-C...H$_2$O complex is found to be more competitive than the one usually proposed (i.e. by successive hydrogenation of s-CO) until $\sim 10^6$ yrs. The introduced reaction path is able to produce a large amount of grain surface and gas-phase methanol. However, in all our models, we still underestimate the gas-phase methanol abundance at least by a factor of $\sim10$ compared with the observations.
\end{enumerate}

\section*{Acknowledgments}
MR, JCL, KMH, PG, FH and VW thanks the following funding agencies for their partial support of this work: the French CNRS/INSU programme PCMI and the ERC Starting Grant (3DICE, grant agreement 336474).

\bibliographystyle{mn2e}
\bibliography{SurfaceReactions_COMs}

\appendix
\section{Reactions added to the network}\label{network}

\begin{table*}
   \begin{center}
   \caption{Grain chemistry}
   \label{grain_chemistry}
   \begin{tabular}{@{}lclcrl}
   \hline
   \hline
   \multicolumn{3}{l}{Reaction} 										& Barrier 	&Branching		&Comments \\
   						&			&						&  (K)	& ratios			&		\\
   \hline
   \multicolumn{6}{c}{Eley-Rideal and complex induced reaction mechanisms} \\
   \hline
   C + s-H$_2$O	 		&$\rightarrow$ & s-C...H$_2$O 			&	0		&				&Ab-initio calculations\\
   C + s-CO$_2$	 		&$\rightarrow$ & s-C...CO$_2$ 			&	0		&				&M06-2X/cc-pVTZ\\
   C + s-NH$_3$			 &$\rightarrow$ & s-C...NH$_3$ 			&	0		&				&Ab-initio calculations\\
   C + s-CH$_4$	 		&$\rightarrow$ & s-C...CH$_4$ 			&	0		&				&Ab-initio calculations\\
   C + s-CH$_3$OH	 		&$\rightarrow$ & s-C...CH$_3$OH 			&	0		&				&Ab-initio calculations\\
   CH + s-H$_2$O	 		&$\rightarrow$ & s-CH...H$_2$O 			&	0		&				&Ab-initio calculations\\
   CH + s-CO$_2$	 		&$\rightarrow$ & s-CH...CO$_2$ 			&	0		&				&Ab-initio calculations\\
   CH + s-NH$_3$	 		&$\rightarrow$ & s-CH...NH$_3$ 			&	0		&				&\citet{Blitz12}\\
   CH + s-CH$_3$OH 		&$\rightarrow$ & s-CH...CH$_3$OH 			&	0		&				&Ab-initio calculations\\
   O + s-CO				&$\rightarrow$ & s-O...CO					&	0		&				&\citet{Goumans10,Talbi06}\\
   C + s-CO				&$\rightarrow$ & s-CCO					&	0		&				&Ab-initio calculations\\
   C + s-H$_2$				&$\rightarrow$ & s-CH$_2$				&	0		&				&\citet{Husain71,Husain75}\\
   						&			&						&			&				&\citet{Harding83,Harding93}\\
   C + s-H$_2$CO			&$\rightarrow$ & s-H$_2$CCO				&	0		&				&\citet{Husain99}\\
   \hline
   \multicolumn{6}{c}{Grain surface reactions} \\ 
   \hline
   s-C...H$_2$O +s-H		&$\rightarrow$ & s-CH$_2$OH				&	0		&	49\%			&	\\
   						&$\rightarrow$ & s-CH$_3$O				&	0		&	49\%			&	\\
   						&$\rightarrow$ & s-CH...H$_2$O			&	0		&	1\%			&	\\
   						&$\rightarrow$ & CH + s-H$_2$O			&	0		&	1\%			&	\\
   s-C...NH$_3$ +s-H			&$\rightarrow$ & s-CH$_2$NH$_2$			&	0		&	98\%			&	\\
   						&$\rightarrow$ & s-CH...NH$_3$			&	0		&	1\%			&	\\
   						&$\rightarrow$ & CH + s-NH$_3$			&	0		&	1\%			&\\
   s-C...CO$_2$ + s-H		&$\rightarrow$ & s-HC(O)CO				&	0		&	0\% 			&\\
      						&$\rightarrow$ & s-HCO + s-CO			&	0		&	98\%			&\\
   						&$\rightarrow$ & s-CH...CO$_2$			&	0		&	1\%			&\\
   						&$\rightarrow$ & CH + s-CO$_2$			&	0		&	1\%			&\\
   s-C...CH$_4$ +s-H			&$\rightarrow$ & s-C$_2$H$_5$			&	0		&				&\citet{Fleurat02}\\
   s-C...CH$_3$OH +s-H		&$\rightarrow$ & s-CH$_3$OCH$_2$		&	0		&	89\%			&\\
   						&$\rightarrow$ & s-CH...CH$_3$OH			&	0		&	10\%			&\\
   						&$\rightarrow$ & CH + s-CH$_3$OH			&	0		&	1\%			&\\
   s-CH...H$_2$O + s-H		&$\rightarrow$ & s-CH$_2$ + s-H$_2$O		& 	0		&				&Ab-initio calculations \\
   s-CH...CO$_2$ + s-H		&$\rightarrow$ & s-CH$_2$...CO$_2$		& 	0		&	 			&\\
   s-CH$_2$...CO$_2$ + s-H	&$\rightarrow$ & s-CH$_3$...CO$_2$		& 	0		&	 			&\\   
   s-CH$_3$...CO$_2$ + s-H	&$\rightarrow$ & s-CH$_4$...CO$_2$		& 	0		&	 			&\\   
   s-CH...NH$_3$ + s-H		&$\rightarrow$ & s-CH$_2$ + s-NH$_3$ 		&	0		&				&\\
   s-CH...CH$_3$OH + s-H	&$\rightarrow$ & s-CH$_2$ + s-CH$_3$OH	&	0		&				&\\
   s-O...CO + s-H			&$\rightarrow$ & s-HOCO					&	0		&	19\%			&\citet{Yu08,Dibble10}\\
						&$\rightarrow$ & s-CO$_2$ + s-H			&	0		&	60\%			&\citet{Yu08,Dibble10}\\
						&$\rightarrow$ & s-OH + s-CO				&	0		&	20\%			&\citet{Yu08,Dibble10}\\
						&$\rightarrow$ & OH + s-CO				&	0		&	1\%			&\\
   s-CH$_3$OCH$_2$ + s-H	&$\rightarrow$ & s-CH$_3$OCH$_3$		&	0		&				&Dimethyl ether formation\\
   s-CH$_2$NH$_2$ + s-H	&$\rightarrow$ & s-CH$_3$NH$_2$			&	0		&				&Ab-initio calculations\\   
   s-C...NH$_3$ + s-N		&$\rightarrow$ & s-HCN + s-NH$_2$		&	0		&				&\citet{Talbi09} \\
   s-HC(O)CO + s-H			&$\rightarrow$ & s-HC(O)CHO				&	0		&				&Glyoxal formation \\
   s-CCO + s-H				&$\rightarrow$ & s-HCCO					&	0		&				&\\
   s-HCCO + s-H			&$\rightarrow$ & s-H$_2$CCO				&	0		&				&\\   
   s-H$_2$CCO + s-H		&$\rightarrow$ & s-CH$_3$CO				&	1720		&				&\citet{Michael79} \\
   s-CH$_3$CO + s-H		&$\rightarrow$ & s-CH$_3$CHO			&	0		&				&\\      
   s-CH$_3$CO + s-CH$_3$	&$\rightarrow$ & s-CH$_3$C(O)CH$_3$		&	0		&				&Acetone formation\\  
   s-HOCO + s-H			&$\rightarrow$ & s-HCOOH				&	0		&	10\% 		&\citet{Yu08,Dibble10}\\
   						&$\rightarrow$ & s-CO$_2$ + s-H$_2$		&	0		&	70\% 		&\citet{Yu08,Dibble10}\\
 						&$\rightarrow$ & s-H$_2$O + s-CO			&	0		&	20\%			&\citet{Yu08,Dibble10}\\
   s-HOCO + s-N			&$\rightarrow$ & s-OH + s-OCN			&	0		&	50\%			&\citet{Dibble10} \\   
   						&$\rightarrow$ & s-NH + s-CO$_2$			&	0		&	50\% 		&\citet{Dibble10}\\
   s-HOCO + s-O			&$\rightarrow$ & s-OH + s-CO$_2$			&	0		&				&\citet{Yu07} \\         
   s-OH + s-CO				&$\rightarrow$ & s-HOCO					&	150		&				&\citet{Fulle96}\\
   						&$\rightarrow$ & s-CO$_2$ + s-H			&	150		&				&\citet{Fulle96}\\
   s-H$_2$CO + s-H			&$\rightarrow$ & s-CH$_3$O				&	2400		&				&\citet{Hippler02} \\
						&$\rightarrow$ & s-CH$_2$OH				&	5400		&				&\citet{Hippler02} \\
						&$\rightarrow$ & s-HCO + s-H$_2$			&	1740		&	 			&\citet{Oehlers00}\\
   s-CH$_3$O + s-H			&$\rightarrow$ & s-CH$_3$OH				&	0		&				&\\
   s-O + s-CH$_3$			&$\rightarrow$ & s-CH$_3$O				&	0		&				&\\
   s-CH$_3$ + s-CH$_3$O	&$\rightarrow$ & s-CH$_3$OCH$_3$		&	0		&				&\\
   s-HCO + s-CH$_3$O		&$\rightarrow$ & s-HCOOCH$_3$			&	0		&				&\\
   \hline
   \end{tabular}
   \end{center}
\end{table*}

\begin{table*}
  \begin{minipage}{180mm}
   \begin{center}
   \caption{Gas phase reactions (temperature range is 10-300K).}
   \label{gas_chemistry}
   \begin{tabular}{@{}lclrrrrrcl}
   \hline
   \hline
   \multicolumn{3}{l}{Reaction} 							&$\alpha$$^a$ 	&$\beta$$^a$&$\gamma$$^a$	&F$_0$$^b$&	g$^c$	&Form.$^d$&	Comments \\
   \hline
   H + CH$_2$NH$_2$ 	&$\rightarrow$& CH$_2$NH + H$_2$		& 3.0(-12)		&	0		&	0		&	3		&	0	&1	&Equal to H + C$_2$H$_5$ \\
   					&$\rightarrow$& CH$_3$ + NH$_2$			& 1.07(-10)	&	0		&	0		&	3		&	0	& 	&\\
   O + CH$_2$NH$_2$ 	&$\rightarrow$& CH$_2$NH + H$_2$		& 1.0(-11)		&	0		&	0		&	2		&	0	&1	&/ O + C$_2$H$_5$ \citet{Harding05}\\
   					&$\rightarrow$& H$_2$CO + NH$_2$		& 1.0(-10)		&	0		&	0		&	2		&	0	&  	&/ O + C$_2$H$_5$ \citet{Tsang86}\\
   C + CH$_2$NH$_2$ 	&$\rightarrow$& C$_2$H$_2$ + NH$_2$		& 3.0(-10)		&	0		&	0		&	3		&	0	&1	&k / C + alkenes and products from \citet{Moskaleva98}\\
   N + CH$_2$NH$_2$ 	&$\rightarrow$& CH$_2$NH + NH			& 4.0(-11)		&	0.17		&	0		&	2		&	7	&1	&/ N + C$_2$H$_5$ \citet{Stief95,Yang05}\\
   					&$\rightarrow$& HCN + NH$_3$			& 6.0(-11)		&	0.17		&	0		&	3		&	21	&	&\\
   H + CH$_2$OH		&$\rightarrow$& H$_2$CO + H$_2$			& 1.0(-11)		&	0		&	0		&	2		&	0	&1	&\citet{Tsang87} \\
   					&$\rightarrow$& CH$_3$ + OH				& 1.6(-10)		&	0		&	0		&	2		&	0	&	&\citet{Tsang87} \\
   O + CH$_2$OH		&$\rightarrow$& H$_2$CO + OH			& 1.0(-10)		&	0		&	0		&	2		&	0	&1	&\citet{Seetula94,Grotheer89}\\
   C + CH$_2$OH		&$\rightarrow$& CH$_3$ + CO				& 3.0(-10)		&	0		&	0		&	3		&	0	&1	&k / C + alkenes and products from \citet{Senosiain06}\\
   N + CH$_2$OH		&$\rightarrow$& H$_2$CO + NH			& 4.0(-11)		&	0.17		&	0		&	2		&	7	&1	&/ N + C$_2$H$_5$ \citet{Stief95,Yang05}\\
   					&$\rightarrow$& HCN + H$_2$O			& 6.0(-11)		&	0.17		&	0		&	3		&	21	&	&\\
   H + CH$_3$O		&$\rightarrow$& H$_2$CO + H$_2$			& 3.0(-11)		&	0		&	0		&	1.6		&	100	&1	&\citet{Hoyermann81,Dobe91} \\
   					&$\rightarrow$& CH$_3$ + OH				& 3.0(-12)		&	0		&	0		&	1.6		&	100	&	&\\
   O + CH$_3$O		&$\rightarrow$& H$_2$CO + OH			& 6.0(-12)		&	0		&	0		&	1.8		&	100	&1	&\citet{Hoyermann81,Dobe91} \\
   					&$\rightarrow$& CH$_3$ + O$_2$			& 1.9(-12)		&	0		&	0		&	1.8		&	100	&	&\\
   C + CH$_3$O		&$\rightarrow$& CH$_3$ + CO				& 3.0(-10)		&	0		&	0		&	3		&	0	&1	&k / C + alkenes and products from \citet{Senosiain06}\\
   N + CH$_3$O		&$\rightarrow$& H$_2$CO + NH			& 1.0(-11)		&	0.17		&	0		&	2		&	7	&	&/ N + C$_2$H$_5$ \citet{Stief95,Yang05}\\
   					&$\rightarrow$& CH$_3$ + NO				& 3.0(-11)		&	0.17		&	0		&	3		&	21	&1	&\\
   CN + CH$_3$OH		&$\rightarrow$& CH$_3$O + HCN			& 6.0(-11)		&	0		&	0		&	3		&	0	&1	&\citet{Sayah98} \\
   					&$\rightarrow$& CH$_2$OH + HCN			& 6.0(-11)		&	0		&	0		&	3		&	0	&	&\citet{Sayah98} \\
   OH + CH$_3$OH		&$\rightarrow$& CH$_3$O + H$_2$O		& 6.0(-13)		&	-1.2		&	0		&	2		&	10	&1	&\citet{Shannon13}\\
   					&$\rightarrow$& CH$_2$OH + H$_2$O		& 3.1(-12)		&	0		&	360		&	1.8		&	100	&	&\citet{Atkinson04}\\
   CH + CH$_3$OH		&$\rightarrow$& CH$_3$ + H$_2$CO		& 2.5(-10)		&	0		&	0		&	1.6		&	10	&1	&Rate constant from \citet{Johnson00}\\
   C + H$_2$CO		&$\rightarrow$& CH$_2$ + CO				& 3.0(-10)		&	0		&	0		&	1.8		&	 0	&1	&\citet{Husain99} \\
   CH + H$_2$CO		&$\rightarrow$& CH$_3$ + CO				& 4.0(-10)		&	0		&	0		&	1.8		&	10	&1	&\citet{Zabarnick88} \\
   CN + H$_2$CO		&$\rightarrow$& HCN + HCO				& 1.0(-11)		&	-0.4		&	0		&	3		&	100	&1	&\citet{Yu93,Chang95} assuming \\
   					&			&						&			&			&			&			&		&	&submerged barrier by comparison with CN + C$_2$H$_6$ \\
   OH + H$_2$CO		&$\rightarrow$& H$_2$O + HCO			& 1.0(-11)		&	-0.6		&	0		&	2		&	10	&1	&\citet{Xu06} \\
   C + CH$_3$OCH$_3$	&$\rightarrow$& CH$_3$O + C$_2$H$_4$	& 3.0(-10)		&	0		&	0		&	3		&	100	&1	&/ C + CH$_3$OH \\
   H + CH$_3$OCH$_2$ 	&$\rightarrow$& CH$_3$O + CH$_3$		& 3.0(-11)		&	0		&	0		&	2		&	100	&1	&/ H + CH$_2$OH \\
   O + CH$_3$OCH$_2$ 	&$\rightarrow$& HCOOCH$_3$ + H			& 2.56(-10)	&	0.15		&	0		&	1.6		&	10	&1	&\citet{Hoyermann96,Song05} \\
   N + CH$_3$OCH$_2$ 	&$\rightarrow$& CH$_3$O + H$_2$CN		& 3.0(-11)		&	0		&	0		&	3		&	10	&1	&/ N + C$_2$H$_5$ \citet{Stief95,Yang05}\\
   C + CH$_3$OCH$_2$ 	&$\rightarrow$& CH$_3$O + C$_2$H$_2$	& 3.0(-10)		&	0		&	0		&	3		&	10	&1	&Capture rate constant, various products possible\\
   C + HCOOCH3		&$\rightarrow$& 2xCO + H + CH$_3$		& 3.0(-10)		&	0		&	0		&	3		&	100	&1	&Capture rate constant, various products possible\\
   					&			&						&			&			&			&			&		&	& but CO is very likely produced\\
   H + HCCO			&$\rightarrow$& CH$_2$ + CO				& 1.7(-10)		&	0		&	0		&	1.6		&	0	&1	&\citet{Glass00,Baulch05}\\
   C + HCCO			&$\rightarrow$& C$_2$H + CO				& 2.0(-10)		&	0		&	0		&	3		&	0	&1	&Estimation \\
   O + HCCO			&$\rightarrow$& H + CO + CO				& 1.6(-10)		&	0		&	0		&	1.6		&	0	&1	&\citet{Baulch05} \\
   					&$\rightarrow$& CH + CO$_2$				& 4.9(-11)		&	0		&	560		&	2		&	100	&	&\\
   N + HCCO			&$\rightarrow$& HCN + CO 				& 6.0(-11)		&	0		&	0		&	3		&	10	&1	&Estimation, part of the HCN lead to \\
   					&$\rightarrow$& HNC + CO 				& 4.0(-11)		&	0		&	0		&	3		&	10	&	&HNC considering the available energy \\
   C + CH$_2$CO		&$\rightarrow$& C$_2$H$_2$ + CO			& 3.0(-10)		&	0		&	0		&	3		&	0	&1	&\\
   C + CH$_3$CHO		&$\rightarrow$& C$_2$H$_4$ + CO			& 3.0(-10)		&	0		&	0		&	1.8		&	0	&1	&\citet{Husain99} \\
   HCCO + H$_3^+$		&$\rightarrow$& H$_2$CCO$^+$ + H$_2$	& 1.0			&	3.1(-9)	&	2.8		&	3		&	0	&2	&Capture rate, this work\\
   HCCO + H$_3$O$^+$	&$\rightarrow$& H$_2$CCO$^+$ + H$_2$O	& 1.0			&	1.4(-9)	&	2.8		&	3		&	0	&2	&Capture rate, this work\\
   HCCO + HCO$^+$		&$\rightarrow$& H$_2$CCO$^+$ + CO		& 1.0			&	1.3(-9)	&	2.8		&	3		&	0	&2	&Capture rate, this work\\
   HCCO + N$_2$H$^+$	&$\rightarrow$& H$_2$CCO$^+$ + N$_2$	& 1.0			&	1.3(-9)	&	2.8		&	3		&	0	&2	&Capture rate, this work\\   
   H$_2$CCO$^+$ + H	&$\rightarrow$& CH$_3^+$ + CO			& 1.0			&	1.94(-9)	&	0.0		&	3		&	0	&3	&Capture rate, this work\\
   CH$_4^+$ + CO		&$\rightarrow$& HCO$^+$ + CH$_3$		& 1.0			&	7.6(-10)	&	0.28		&	2		&	0	&3	&\citet{Anicich98} \\
   \hline
   \end{tabular}
   \end{center}
   \medskip
   $^a$$\alpha$, $\beta$ and $\gamma$ are the parameters used to compute the rate of the reaction.\\
   $^b$F$_0$ is an uncertainty parameter.\\
   $^c$$g$ is used to parametrize a possible temperature-dependence of the uncertainty F$_0$.\\
   $^d$Formula ton compute the temperature dependent rate coefficient: 1: k(T$_\textrm{gas}$)$_\textrm{Kooij}$ = $\alpha$(T$_\textrm{gas}$/300)$^\beta$e$^{-\gamma/\textrm{T}_\textrm{gas}}$, 2: k(T$_\textrm{gas}$)$_\textrm{Ionpol1}$=$\alpha \beta(0.62 + 0.4767 \gamma (300/\textrm{T}_\textrm{gas})^{0.5})$ 3: k(T$_\textrm{gas}$)$_\textrm{Ionpol2}$=$\alpha \beta(1.0 + 0.0967 \gamma (300/\textrm{T}_\textrm{gas})^{0.5} + (\gamma^2/10.526)(300/\textrm{T}_\textrm{gas}))$\citep[see][for more information]{Wakelam12}.
\end{minipage}
\end{table*}

\end{document}